# Understanding and Tuning Magnetism in van der Waals-type Metal Thiophosphates

Rabindra Basnet[1*] and Jin Hu[2,3#]

[1]Department of Chemistry & Physics, University of Arkansas at Pine Bluff, Pine Bluff, AR, 71603 USA

[2]Department of Physics, University of Arkansas, Fayetteville, Arkansas 72701, USA

[3]Materials Science and Engineering Program, Institute for Nanoscience and Engineering, University of Arkansas, Fayetteville, Arkansas 72701, USA

Corresponding author: *basnetr@uapb.edu; #jinhu@uark.edu

Abstract

Over the past two decades, significant progress in two-dimensional (2D) materials has invigorated research in condensed matter and material physics in low dimensions. While traditionally studied in three-dimensional systems, magnetism has now been extended to the 2D realm. Recent breakthroughs in 2D magnetism have captured substantial interest from the scientific community, owing to the stable magnetic order achievable in atomically thin layers of the van der Waals (vdW)-type layered magnetic materials. These advances offer an exciting platform for investigating related phenomena in low dimensions and hold promise for spintronic applications. Consequently, vdW magnetic materials with tunable magnetism have attracted significant attention. Specifically, antiferromagnetic metal thiophosphates $M\mathrm{P}X_3$ ($M$ = transition metal, $P$ = phosphorus, $X$ = chalcogen) have been investigated extensively. These materials exhibit long-range magnetic orders spanning from bulk to the 2D limit. The magnetism in $M\mathrm{P}X_3$ arises from localized moments associated with transition metal ions, making it tunable via substitutions and intercalations. In this review, we focus on such tuning by providing a comprehensive summary of various metal- and chalcogen-substitution and intercalation studies, along with the mechanism of magnetism modulation, and a perspective on the development of this emergent material family.

# Table of Content

## 1. Introduction

Since the discovery of graphene[1], numerous two-dimensional (2D) materials have been theoretically predicted and experimentally realized. For a long time, 2D material research was mainly centralized around graphene and transition metal dichalcogenides (TMDs) family in terms of their electronic properties because they display almost all functionalities of solid-state systems such as metal, semimetal, semiconductor, and insulator as well as exotic quantum phases and phenomena such as superconductivity, charge density wave, and topological quantum phases. In addition to these properties, magnetism in 2D materials has also attracted intensive attention due to fundamental scientific interest in low dimensional orders and technological demands such as spintronics. Graphene[2] and TMDs[3–5] have been found to exhibit magnetism. However, owing to the non/weak-magnetic nature of the constituted atoms/ions, magnetism in these materials mainly arises from extrinsic factors such as defects or doping of guest atoms. Therefore, there has been an intensive search for 2D materials with intrinsic magnetism. Recently, there have been breakthroughs in discovering intrinsic magnetic orders in atomically thin layers of $FePS_3$[6], $CrI_3$[7], $CrGeTe_3$[8], $CrBr_3$[9], $VI_3$[10], $Fe_3GeTe_2$[11] and $Fe_3GaTe_2$[12], which greatly expand the realm of the 2D magnetism and provide material platforms not only for gaining deep insight into low dimensional magnetism but also for spintronic applications[13–25]. These developments have motivated the search for additional novel magnetic van der Waals (vdW) materials, resulting in a blossom of magnetic vdW materials that can be broadly categorized into two groups: (1) transition metal halides (including both dihalides and trihalides), and (2) transition metal chalcogenides (both binary and ternary) involving transition metals from V to Ni in the periodic table (i.e., V, Cr, Mn, Fe, Co, and Ni). Magnetism in those vdW magnets arises from the magnetic moment of the 3*d* electrons of the transition metal cations. The metal ions in these compounds are mostly arranged in honeycomb lattices, except for a few compounds such as transition metal dihalides that is characterized by triangular networks of metal ions[26].

Metal thiophosphates are one of the model systems among various 2D magnetic materials. Those materials were discovered in the late 1800s by Friedel and Ferrand[27,28], usually represented by a chemical formula of $MPX_3$ or $M_2P_2X_6$, where $M$, P, and X denote transition metal, phosphorus, and chalcogen elements, respectively. Since their discovery, $MPX_3$ compounds have been studied in various fields including magnetism[6,29–43], magneto-optics[44–50], optoelectronics[27,28,51–53], Li-ion batteries[27,28,54,55], and ferroelectricity[27,28,56–59]. Their layered structures allow for mechanical exfoliations down to atomically thin layers[6,29,41–43,60], creating pathways to study 2D physics. The magnetic members of the $MPX_3$ ($M$ = V, Cr, Mn, Fe, Co, and Ni) family are antiferromagnets in which the $M$ ions carry localized magnetic moments in a layered honeycomb lattice. Antiferromagnetic (AFM) orders have been found to persist down to atomically thin layers and even to the monolayer limit in some $MPX_3$[6,42,43], thus these materials have been established as important candidates for 2D magnets[6,29,41–43,60]. The $MPX_3$ compounds exhibit semiconducting[61] or insulating[62,63] behavior with band gaps ranging from 1.3 to 3.5 eV. Interestingly, applying high pressure can induce insulator-to-metal transitions[64–71] and even superconductivity[72] in some $MPX_3$ compounds. Furthermore, these compounds also offer platforms to study correlated electrons in 2D magnetic materials[64,73,74]. Recent studies have revealed a strong coupling between optically active excitons and magnetism in $MPX_3$[44–49]. For example, in $NiPS_3$, a novel coherent spin-orbit entangled excitonic state stabilized by AFM order[44] as well as the correlation between emitted photon and spins[45] have been reported. Such coupling can lead to an all-optical control of magnetic anisotropy[50]. In addition, a strong correlation of magnons and phonons that generates phonon spin through the transfer of spin angular momentum has been demonstrated in $FePS_3$[42,75]. Such intertwined degrees of freedom provide a unique opportunity to explore the interplay between magnetism, strong correlation, and light-matter interactions in 2D magnets.

In addition to fundamental research, this material family also offers a large material pool for potential technological applications. The feasibility of obtaining atomically thin layers enables nanodevices and heterostructures fabrication and integration, opening opportunities for utilization in electronics and spintronics. Moreover, in $MPX_3$, the band gaps ranging from 1.3 to 3.5 eV cover near-infrared to the UV region, enabling the optoelectronic applications in a broad wavelength horizon[27,28]. Additionally, the unusual intercalation-reduction behavior and higher ionic conductivity of these materials facilitate potential applications in Li-ion batteries[27,28,54,55]. Furthermore, their unusual ferroelectricity could lead to new energy and data storage devices and sensors[27,28].

Given such versatility, $MPX_3$ materials have recently gained intensive attention. In particular, magnetism in those materials has been established as an important research topic[76]. The magnetic properties in $MPX_3$ strongly depend on the choice of the transition metal $M$ and chalcogen $X$ elements. Not only the AFM transition temperature (Néel temperature, $T_N$) that changes upon substituting $M$ and $X$[30–39], but also various AFM structures can be obtained through $M$[77–94] and $X$[30–32,95,96] substitutions. Such tuning of magnetism by altering $M$ and $X$ has led to the study of a series of polymetallic[77–94] and polychalcogenide[30–32,95,96] "mixed" $MPX_3$ compounds. Tunable magnetism has been observed in these mixed compounds, providing promising candidates to explore novel phenomena originating from 2D magnetism. In addition to substitution, the layered structure of $MPX_3$ allows for another doping strategy i.e., the inter-layer intercalation of guest ions. Intercalating various guest species has been reported to cause AFM to ferrimagnetic (FIM)[97–101] and ferromagnetic (FM)[102] transitions in $MPX_3$. This suggests that the intercalation could be a promising route to achieve additional magnetic phases in $MPX_3$ that may not be accessible by substitution.

Thus far, a wide range of magnetic, electronic, vibrational, optical, and ferroelectric properties of pristine $MPX_3$ compounds have already been summarized by a few excellent review articles[27,28,76]. Therefore, the focus of this review will be put on engineering magnetic properties through various doping strategies: metal substitution, chalcogen substitution, and inter-layer intercalation, aiming to provide a comprehensive summary with sufficient depth and width for deeper understanding of magnetism in $MPX_3$ based on magnetic exchanges and anisotropies. This review is organized as follows: summarizing the crystal growth methods and lattice structures, an overall theoretical framework for understanding magnetism in $MPX_3$, experimental efforts in engineering the magnetism, and the perspective of the development of this field. We hope this review can provide the community with comprehensive information on the existing discoveries in doping studies and an in-depth understanding of magnetism, which serves as an important guideline for further exploring the exciting materials physics in $MPX_3$ as well as other 2D magnets and their property tuning. We wish to cover all relevant studies in this review. However, given the rapid development of $MPX_3$ materials, the emerging new discoveries may not be included but can be introduced by future review articles.

## 2. Crystal growth

The studies on functional materials, their properties, and applications, greatly rely on the development of convenient and reliable synthesis methods to produce high-quality crystals. For solid state systems, single crystals are highly desired because a single crystal possesses well-defined crystalline axes so that the intrinsic anisotropic magnetic, electronic transport and optical properties can be experimentally determined, which are critical to understanding the fundamental physics principles that govern the material properties. In addition, well-defined surfaces and edges of a single crystal enables the surface-sensitive measurements such as angle resolved photoemission spectroscopy (ARPES), scanning tunneling microscopy/spectroscopy (STM/S), etc. which provide valuable information. Conventional single-crystal synthesis techniques include chemical vapor transport (CVT), flux, Bridgman, floating zone (FZ), and Czocharalski methods. The synthesis of $MPX_3$ compounds goes back to the late 19$^{th}$ century when Friedel obtained $FePS_3$ by heating a weighed amount of phosphorus pentasulfide with iron[27,28]. This technique was later used by Ferrand[27,28] to synthesize other sibling compounds $ZnPS_3$, $CdPS_3$, and $NiPS_3$. Nowadays, CVT and flux have become the main techniques to grow the bulk single crystals of $MPX_3$, which have proven to be effective methods to yield large single crystals for physical characterizations.

A CVT method has been widely used to synthesize pristine, metal-, and chalcogen-substituted $MPX_3$ compounds as summarized in Table 1. As shown in Fig. 1(a), this method is based on the vaporization of source materials at the hot end of the growth chamber (usually a quartz ampoule), transportation of the vapor along some temperature gradient, and condensation of vapor at the cold end. For $MPX_3$, a mixture of elementary powders of $M$, P, and $X$, or pre-synthesized polycrystal precursors prepared by solid-state reaction, can be used as source materials for CVT. For CVT growth, a transport agent such as $I_2$, $TeCl_4$, $SeCl_4$, etc., is usually added to the source material to aid the vapor transportation process. For $MPX_3$, CVT growth using these transport agents has been reported[31,40,71,88–92,103–107]. On the other hand, owing to the volatile elements P and S or Se, CVT growth without a transport agent[33,41,108–113] has also been successfully used to synthesize the single crystals of $MPX_3$. Experimentally, the source materials (elementary powders with proper molar ratio or pre-reacted polycrystal precursors) along with the transport agent (if used) are sealed in an evacuated quartz ampoule. The ampoule should be sufficiently long (usually ~10-15 cm) to create a proper temperature gradient between the source and sink to facilitate the vapor transport. Then, the ampoule is placed in a two-zone furnace [Fig. 1(a)] which provides the needed temperatures and temperature gradient and heated at a certain duration required for the formation of the desired $MPX_3$ phase. The source end is kept at a higher temperature ($T_{hot}$) than the sink end ($T_{cold}$) to establish the thermal gradient, as summarized in Table 1. The heating process must be slow to avoid excessive pressure due to volatile elements P and $X$. Upon the completion of the growth, the furnace either slowly cools at a controlled slow cooling rate or naturally cools to room temperature, and single crystals can be found at the cold end. Single crystals harvested by quenching in ice water (i.e., very rapid cooling) have also been reported[114]. The optical images of single crystals of some pristine and substituted (both metal- and chalcogen-substituted) $MPX_3$ compounds synthesized using the CVT technique are shown in Figs. 1(b-d).

In addition to CVT, the reactive flux technique has been recently reported to synthesize pristine and metal-substituted $MPX_3$ single crystals. This method adopts $P_2X_5$ ($X$ = S[115–119] or Se[120]) flux and metal precursors, as summarized in Table 1. First, the metal and $P_2X_5$ powders in a ratio of 2:3 for monometallic or 1:1:3 for bimetallic compounds is sealed in an evacuated silica tube inside a glove box. The mixture is then heated and maintained at a certain temperature with different heating profiles depending on the target compounds[116]. The excess $P_2S_5$ (melting point ~285 °C)[115–119] and $P_2Se_5$ (melting point ~214 °C)[120] act as a reactive flux that oxidizes the metal element and enhances solubility to reduce reaction times and increase crystallite sizes. Upon the completion of the growth, the excess $P_2S_5$ flux is then removed with a 1:1 mixture of ethanol and water followed by a subsequent treatment using deionized water and acetone[115–119]. On the other hand, the excess $P_2Se_5$ flux is removed by the distillation process at a temperature gradient between 400 °C and room temperature[120]. This method can result in sizable single crystals in a relatively shorter synthesis time as compared to a traditional CVT technique, as shown in Fig. 1(e).

**3. Crystal structure**

The crystal lattice structures and symmetries are critical in determining magnetic properties, as has been theoretically predicted for $MPX_3$[121]. Varying lattice parameters have been found to strongly affect the magnetic properties[121]. So, understanding crystal structure is essential for gaining deeper insights into magnetism. At first glance, all $MPX_3$ materials exhibit very similar layered structures [Fig. 2(a)]. Careful structure characterizations have revealed monoclinic, rhombohedral, and triclinic structures for various $MPX_3$, as summarized in Table 2. Despite three different lattice structures, $MPX_3$ materials exhibit common structural characteristics. As shown in Fig. 2(b), metal atoms $M$ are arranged in a honeycomb lattice. Each P displays a tetrahedral coordination, which is covalently bonded with one P to form P-P ($P_2$) dimers perpendicular to the hexagonal plane and three $X$ featuring a $(P_2X_6)^{4-}$ bipyramid structure unit. Such P-P dimer forming $(P_2X_6)^{4-}$ bipyramids fill the center of the $M$ honeycomb lattice. Therefore, these compounds have also been referred to as $M_2P_2X_6$ in the literature. For chalcogen atoms $X$ that sandwich the metal honeycomb layer, they form almost close-packed surfaces of the layer with an array of octahedral coordinated sites [Fig. 2(c)]. The monolayer $MPX_3$ can be considered as the monolayer $MX_2$ where the one-third of $M$ sites are substituted by P-P dimers. Thus, in $MPX_3$, two-thirds of the octahedral centers are filled by $M^{2+}$ cations while the P-P dimers occupy the remaining one-third [Fig. 2(c), P-P dimer in the center is not shown].

As seen in Fig. 2(a), a finite vdW gap of about 3.22-3.24 Å is present in $MPX_3$[27,28,76]. Various metal ions $M^{2+}$ with different ionic radii slightly modify the slab size. The two flat pyramids formed by bonding between a P and three $X$ atoms at the top and bottom of the $(P_2X_6)^{4-}$ structural unit [shown in Fig. 2(b)] remain invariable, but the P-P distance is slightly adjusted to accommodate the changes in metal cations. For example, the P-P distance elongates from 2.148 Å in $NiPS_3$ to 2.222 Å in $CdPS_3$[27], which is accompanied by increases of the layer thickness.

Although the individual lamella is similar for all $MPX_3$ compounds, their symmetry, and layer stacking to form a bulk lattice structure can vary depending on $M$ and $X$, as summarized in Table 2. In particular, the symmetries and crystal structures for sulfides and selenides of $MPX_3$ materials are distinct. Generally, almost all the known sulfides $MPS_3$ exhibit monoclinic crystal structures with a $C2/m$ space group, except for $PbPS_3$[106] which has been reported to crystalize in a $P21/c$ monoclinic structure as its selenide counterpart, and for $HgPS_3$[119] which is triclinic with a $P\bar{1}$ space group symmetry. In sulfides, the monoclinic angle $\beta$ depends on the metal cation, ranging from 106.97° for $MgPS_3$ to 107.35° for $MnPS_3$. On the other hand, as summarized in Table 2, selenides $MPSe_3$ process larger P-Se bond lengths and Se-P-Se bond angles as compared to their sulfide counterparts, and mostly crystallize in rhombohedral structures with a $R\bar{3}$ space group. The exceptions are $NiPSe_3$, $CrPSe_3$, $HgPSe_3$, and $PbPSe_3$ which exhibit monoclinic crystal structures.

Therefore, the lattice structure of $MPX_3$ appears to be more sensitive to chalcogen rather than metal. This is further confirmed by substitution studies. As summarized in Table 2, substituting metal atoms $M$ does not change the lattice structure in $MPX_3$[77–93,115,117,118]. Most known metal-substituted sulfide compounds exhibit monoclinic structure with a $C2/m$ space group in the entire substitutional range. For selenides, previous works on a bimetallic compound $Mn_{1-x}Fe_xPSe_3$[87] and a high entropy compound $Fe_{1-x-y-z}Mn_xCd_yIn_zSe_3$[122] also found unchanged lattice with rhombohedral $R\bar{3}$ structures. On the other hand, chalcogen S-Se substitution can induce structure transition from monoclinic ($C2/m$) to rhombohedral ($R\bar{3}$) in $MnPS_{3-x}Se_x$[32] and $FePS_{3-x}Se_x$[123], which is consistent with the different lattice structures for their sulfide and selenide end compounds. One exception is $NiPS_{3-x}Se_x$ which displays unchanged monoclinic *C*2/*m* structure, which is also in line with the identical *C*2/*m* space group for $NiPS_3$[31,89,90,92,124] and $NiPSe_3$[125].

# 4. Magnetism

## 4.1. Theoretical background

Owing to the insulating nature of $MPX_3$ compounds, their magnetism can be described by local moment pictures. Magnetism in vdW magnetic materials arises from the magnetic moment of the spin and orbital momenta of the 3*d* electrons of the transition metal cations. Thus, the choice of transition metal cations plays a vital role in determining magnetism. In the presence of a crystal electric field from the octahedral coordination of the transition metal cations [Fig. 2(c)], the $d_{x^2-y^2}$ and $d_{z^2}$ orbitals among the five *d*-orbitals of the transition metal point in the direction of the ligands, thereby the electrons in these orbitals experience greater repulsion with ligands and increase the energy of these orbitals. Thus, these two *d*-orbitals are pushed to higher energy levels in the band structure and referred to as two-fold degenerate high-energy $e_g$ orbitals. On the other hand, the remaining three *d*-orbitals $d_{xz}$, $d_{yz}$, and $d_{xy}$ do not directly align towards ligands leading to a weaker repulsion and stabilized to a lower energy level than two $e_g$ orbitals. Therefore, these three orbitals lie below $e_g$ orbitals ($d_{x^2-y^2}$ and $d_{z^2}$) in term of energy and constitute three-fold degenerate low-energy $t_{2g}$ orbitals[126,127]. The splitting of five 3*d* orbitals of transition metal ions under an octahedral crystal field is depicted in the schematic in Fig. 2(d). Then, the *d* electrons of various transition metal cations fill these orbitals following Hund's rule, as presented in Table 3[128]. The occupancy of these orbitals determines the total magnetic moment of the 3*d* transition metal cations, with the total magnetic moment given by $g_J[j(j+1)]^{1/2}\mu_B$ where $|j|$, $g_J$, and $\mu_B$ are the total angular momentum quantum number (= spin + orbital angular momentum), Landé *g*-factor, and Bohr magneton, respectively. When orbital angular momentum is quenched (i.e., $L = 0$), only the spin magnetic moment contributes to the total magnetic moment, resulting in a total magnetic moment of $2[s(s+1)]^{1/2}\mu_B$ where $s$ is the total spin quantum number, which is generally the case for transition metals[126].

In the classic Heisenberg model, the magnetism can be explained within the framework of isotropic Heisenberg Hamiltonian: $H = -2\Sigma(JS_i \cdot S_j)$, where $J$ is the exchange coupling, which is considered to be isotropic, the summation is over all pairs of magnetic ions in the lattice and the spins are treated as three-component vectors where $S_i$ and $S_j$ are spin magnetic moments of the atomic site $i$ and $j$ respectively. The Heisenberg model provides an accurate account of the interactions between the localized magnetic moments that govern the magnetic properties in insulators (such as $MPX_3$). In vdW materials, magnetic interactions can be anisotropic and are described by a Heisenberg Hamiltonian with anisotropic exchange: $H = -2\Sigma(J_xS_iS_j + J_yS_iS_j + J_zS_iS_j)$, where $J_x$, $J_y$, and $J_z$ are exchange couplings along the crystallographic axes. In layered $MPX_3$, magnetic exchanges within the layer are much stronger than inter-layer interactions[129–131], hence the long-range magnetic ordering is believed to be mainly governed by intralayer exchange interactions. The interactions between localized moments in insulating $MPX_3$ compounds are mainly mediated through direct and superexchange interactions, as shown in Fig. 2(e). The direct exchanges between neighboring magnetic moments, which result from the overlapping wavefunctions, mainly depend on their spatial separation, and contribute to the nearest neighbor interaction $J_1$. Despite the fact that the distances between magnetic cations are large (in the order of 3-4 Å), such direct exchange has been found to be important or even dominate the magnetic interactions in some $MPX_3$ compounds such as $MnPS_3$[129]. On the other hand, in many other $MPX_3$ materials such as $NiPS_3$, the long-range magnetic coupling between moments is dominated by metal-chalcogen-metal ($M$-$X$-$M$) superexchange interactions mediated through non-magnetic chalcogen $X$[129]. The superexchange couplings contribute to both the nearest-neighbor interaction ($J_1$) and the interactions between further neighbors such as second nearest-neighbor ($J_2$), and third nearest-neighbor ($J_3$) interactions[129]. The magnetic interactions $J_1$, $J_2$, and $J_3$ are depicted in Fig. 2(f). The nature and strength of superexchange interactions are determined by $M$-$X$-$M$ bonding angle, $d$-orbital occupancy of $M^{2+}$ ions, and orbital overlap between $M^{2+}$ cation and $X^{2-}$ anion[40,129]: According to Goodenough-Kanamori rules[132,133], the superexchange interaction is ferromagnetic (FM) if $M$-$X$-$M$ bond angle is close to 90° and AFM when $M$-$X$-$M$ bond angle is 180°. The occupancy of $d$ orbitals is important. For example, $CoPS_3$ and $NiPS_3$ share the same magnetic structure but possess different strengths for superexchange due to their distinct occupancy of $d$ orbitals for $Co^{2+}$ ($d^7$) and $Ni^{2+}$ ($d^8$) ions[130]. Such differences could be important in designing different strategies to tune magnetism in $CoPS_3$ and $NiPS_3$. The overlap of atomic orbitals also affects the superexchange interaction. Usually, stronger orbital overlap strengthens the superexchange, therefore substituting Se for S in $MPX_3$ leads to enhanced superexchange because of the extended orbital overlap due to more extended atom orbitals of Se than S[31].

Furthermore, structural distortions also affect the magnetism in $MPX_3$[40,129,130]. As illustrated in Fig. 2(c) and mentioned above, every metal atom $M$ in $MPX_3$ is located at the center of an octahedron formed by six $X$ atoms. Such $MX_6$ octahedra, however, possess a trigonal distortion that is characterized by the angle $\theta$ between the trigonal axis (perpendicular to the $ab$-plane) and the $M$-$X$ bond [Fig. 2(c)]. The experimental $\theta$ values for various $MPX_3$ compounds are presented in Table 4. An ideal octahedral possesses the ideal octahedral angle $\theta \approx 54.75°$. The deviation of $\theta$ angle from this ideal value characterizes the trigonal distortion. The crystal field arising from trigonal distortion of $MX_6$ octahedra together with spin-orbit coupling (SOC) of the $M^{2+}$ cations introduce anisotropy in an otherwise isotropic system. Now, with an additional anisotropy term, the full Hamiltonian becomes $H = -2\ \Sigma(J_{ij}S_i \cdot S_j) - A\ \Sigma(S_i^z)^2$, where $J_{ij}$ is exchange interaction between neighboring $i^{th}$ and $j^{th}$ moments and $A$ is single-ion anisotropy parameter that defines the magnetic anisotropy of the system. The strengths of various *J's* and $A$ determine the magnetism in $MPX_3$. Table 4 summarizes the experimental values of $J$ and $A$ parameters obtained by neutron scattering experiments for various $MPX_3$ materials[35,95,108,129,130], together with theoretical calculated values for $VPS_3$[134]. These distinct $J$ and $A$ values for different $MPX_3$ compounds result in diverse AFM structures for this material family, as shown in Fig. 3. When $J \gg A$ (for very small $A$), the magnetism is weakly anisotropic and can be explained by the 3D isotropic Heisenberg Hamiltonian, such as $MnPS_3$[35,40]. For large $A$ values, on the other hand, the magnetism becomes rather anisotropic and corresponds to *XY* (such as $NiPS_3$[40,124] and $CoPS_3$[130]) or Ising-type (such as $FePS_3$[40] and $FePSe_3$[30]) magnetism depending on the sign and magnitudes of $A$. In Mn and Ni systems, the orbital angular momentum ($L$) is quenched for $Mn^{2+}$ ($d^5$) and $Ni^{2+}$ ($d^8$) ions because of their half and fully-filled $t_{2g}$ orbitals respectively, leading to negligible SOC for magnetic ions[40]. In $MnPS_3$ which possesses relative weak trigonal distortion (octahedral angle $\theta \approx 51.67°$), the effect of both spin-orbit splitting and trigonal distortion is found to be negligible for the high spin ground state of $Mn^{2+}$[40], so the magnetism in $MnPS_3$ is governed by the dipolar anisotropy[35,135,136] which results in an out-of-plane moment orientation with a small tilt towards the *a*-axis (Fig. 3). On the other hand, $NiPS_3$ exhibits greater trigonal distortion characterized by octahedral angle, $\theta \approx 51.05°$ from the undistorted octahedral angle of 54.75°. This leads to remarkable single-ion anisotropy $A \approx 0.3$ meV in $NiPS_3$, which is much stronger than that of 0.0086 meV in $MnPS_3$ (Table 4)[35,129,130]. Consequently, Ni moments are aligned within the basal plane perpendicular to the trigonal axis (Fig. 3)[36,40,136], and this system can be modeled by the anisotropic

Heisenberg Hamiltonian with *XY*-type anisotropy. The situation is different for $FePS_3$, which exhibits a substantial magnetic anisotropy with a much higher $A$ ($\approx$ 2.66 meV) in comparison to $MnPS_3$ and $NiPS_3$[108,129]. Thus, $FePS_3$ is best described by the Ising Hamiltonian corresponding to the highly anisotropic Ising-type magnetic ordering with magnetic moments aligned along the out-of-plane direction[40,108]. In $FePS_3$, the degenerate $t_{2g}$ orbitals for the high-spin $Fe^{2+}$ ($d^6$) state lead to a stronger trigonal distortion of the $FeS_6$ octahedra ($\theta \approx 51.28°$[108,130]) that causes a significant splitting of $t_{2g}$ orbitals to lift the degeneracy while the $e_g$ orbitals remain doubly degenerate[137]. In addition, the lack of half- or fully-filled $t_{2g}$ orbitals for high-spin $Fe^{2+}$ ions also contribute to the net orbital moment and consequently enhances the SOC. Although, a recent x-ray photoemission electron microscopy study has demonstrated that the spin-orbit entanglement for $Fe^{2+}$ under trigonal elongation plays a key role behind the significant magnetic anisotropy in $FePS_3$[138], the SOC alone may not be a dominant factor behind strong magnetic anisotropy in $FePS_3$ because the SOC for $Ni^{2+}$ ($d^8$) ion is larger than $Fe^{2+}$ ($d^6$) ion but the resulting compound $NiPS_3$ is less anisotropic than $FePS_3$[40]. The large crystal electric field arising from the strong trigonal distortion of the $FeS_6$ octahedra might also be attributed to a substantial magnetic anisotropy in $FePS_3$, which has been confirmed by a recent x-ray absorption spectroscopy experiment[137]. In fact, the strong anisotropy due to crystal electrical field i.e., crystal-field anisotropy has been ascribed to $Fe^{2+}$ ($d^6$) ion in different Fe-based compounds[139,140].

### 4.2. Engineering magnetism in $\boldsymbol{MPX_3}$

Engineering magnetism in layered magnetic materials not only optimizes properties for technological applications (such as high transition temperature, wider hysteresis loop, and various types of FM and AFM structures) but also provides an effective approach to clarify the mechanism for the magnetism by identifying the parameters that govern magnetism. Several strategies such as doping[30–32,77–92,95,96,98–100,141–147], high pressure[148–151], and electrostatic gating[14,16] have been adopted to tune the magnetic properties in layered magnets. These techniques have been successful in controlling the spin orientation in vdW magnetic materials, which leads to novel magnetic phenomenon arising from 2D magnetism. Tuning magnetic properties has been one major focus and studied for a long time in the $MPX_3$ family[78,79,86]. Doping, including metal $M$ and chalcogen $X$ substitutions, and inter-layer intercalation, has been demonstrated as a powerful technique to modulate magnetism in $MPX_3$, which will be the focus of this review as discussed below.

### 4.2.1. Metal (*M*) substitution

Metal substitution has been adopted as an important method to manipulate the magnetic properties in $MPX_3$. Owing to the structural similarity of the $MPX_3$ materials, a divalent metal $M^{2+}$ ion can be substituted with another divalent metal ion to create a polymetallic "mixed" compound such as $(Mn,Ni)_xFe_{1-x}PS_3$, $Ni_{1-x}Mn_xPS_3$, $(Mn,Fe)_{1-x}Zn_xPS_3$, $(Mn,Fe,Ni)_{1-x}Co_xPS_3$ and $Mn_{1-x}Fe_xPSe_3$[77–92,94,115,117,118,152,153] where two different metal atoms are mixed in a honeycomb lattice as shown in Fig. 4(a). In addition to these bimetallic substitutions, the metal substitution in $MPX_3$ has recently been extended to mixing of more than two divalent metal $M^{2+}$ ions creating medium[122] and high entropy[114] vdW materials. In addition to these bi- and poly-metallic compounds based on isovalent $M^{2+}$ ion substitutions, the non-isovalent substitution of monovalent and trivalent ions that leads to a series of quaternary compounds $(M_1)^{1+}(M_2)^{3+}P_2X_6$ ($M_1$ = Li, Cu, Ag, etc; $M_2$ = V, Cr, In, Bi, etc; $X$ = S and Se)[120,154–162] have also been reported in this material family. Although the focus of this review is the divalent metal substitutions, we have also briefly summarized the magnetic properties of a few magnetic quaternary compounds $(Cu,Ag)^{1+}(V,Cr)^{3+}P_2(S,Se)_6$ that have been formed by mixing trivalent ions such as $Cr^{3+}$ or $V^{3+}$ with monovalent metal ions such as $Cu^{1+}$ or $Ag^{1+}$[154–158]. The magnetic exchange $J$ and single-ion anisotropy $A$ parameters are found to be effectively modulated with metal substitutions that lead to efficient control of magnetic properties. Tunable magnetism arising from competing magnetic interactions, magnetic anisotropies, and spin fluctuations has been observed in metal-substituted $MPX_3$ compounds, which will be discussed in the following sections.

#### 4.2.1.1. Isovalent substitution

##### 4.2.1.1.1. $Mn_{1-x}Fe_xPS_3$

As shown in Fig. 3, $MnPS_3$ and $FePS_3$ exhibit out-of-plane magnetic moment but different Néel[135] and zig-zag[108] type AFM structures for each layer, respectively. Therefore, the tuning of magnetic properties arising from the competition between two different AFM interactions is expected when $Mn^{2+}$ and $Fe^{2+}$ ions are mixed, which has been studied by Masbuchi et al.[82] Single crystals of $Mn_{1-x}Fe_xPS_3$ ($0 \leq x \leq 1$) were synthesized using a CVT technique as detailed in Ref. [82].

As shown in Fig. 5, the magnetic properties were characterized by the temperature dependence of magnetization under various magnetic fields along the easy axis (the $c*$ axis in Fig. 5 is perpendicular to the $ab$-plane. i.e., the out-of-plane direction)[82]. A few magnetic phases have been identified: AFM order for $0 \leq x \leq 0.2$ and $0.8 \leq x \leq 1$ composition, which is characterized by a clear AFM transition in susceptibility without remarkable difference between zero-field-cooling (ZFC) and field-cooling (FC). A broad hump in susceptibility occurs above the AFM ordering temperature $T_N$ in end compounds [Fig. 5(a)]. This is a generic feature in many $MPX_3$ compounds, which has been attributed to a short-range 2D or quasi-2D magnetic correlation owing to the layered structures. The broad hump is suppressed by substitution, leading to a sharper peak at $T_N$ [Fig. 5(a)], which has been ascribed to the weakening of a short-range magnetic correlation due to randomness introduced by Fe substitution.

Samples in $0.5 \leq x \leq 0.6$ compositions display significant irreversibility [Fig. 5(b)] and are ascribed to a spin glass phase, which might be attributed to the competition between dominant exchange interactions in $MnPS_3$ and $FePS_3$ phases, namely AFM $Mn^{2+}$-$Mn^{2+}$ $J_1$ interaction and FM $Fe^{2+}$-$Fe^{2+}$ $J_1$ interaction, respectively[35,108] (Table 4). This is further confirmed by AC susceptibility[82]. The glass temperature, defined as the temperature where susceptibility displays irreversibility between ZFC and FC measurement, is found to enhance with the frequency in AC susceptibility measurements. The remaining composition regions, i.e., $0.3 \leq x \leq 0.45$ and $0.6 < x \leq 0.7$, are characterized by an AFM-like transition without irreversibility at $T_N$. However, irreversibility occurs at a lower temperature. This has been ascribed by a reentrant spin glass transition [RSG in the phase diagram in Fig. 5(c)]. Considering equal Mn-Fe substitutions in both end compounds, the reentrant spin glass phase is found to be stronger towards Fe-rich compositions as compared to Mn-rich compositions, which is manifested by the higher reentrant spin glass transition temperature for $x = 0.7$ than $x = 0.3$[82]. Such behavior could stem from much larger single-ion anisotropy $A$ for $FePS_3$ (Table 4) as compared to $MnPS_3$, as has been proposed in dilute alloys[163].

Overall, $Mn_{1-x}Fe_xPS_3$ displays a non-monotonic doping dependence of $T_N$ with a minimum of around $x \approx 0$. Such behavior is widely seen in other metal-substituted $MPX_3$ and is usually interpreted by spin frustration due to the mixing of two types of magnetic ions compounds ($MnPS_3$ and $FePS_3$) with significantly distinct magnetic exchange interactions $J_1$, $J_2$, and $J_3$[35,108] as summarized in Table 4.

#### 4.2.1.1.2. $Fe_{1-x}Ni_xPS_3$

For $Fe_{1-x}Ni_xPS_3$, the distinct *XY*- and Ising-like antiferromagnetism for the end members $NiPS_3$ and $FePS_3$ respectively makes it rather interesting to investigate the evolution between those two magnetic orders with distinct magnetic anisotropies.

$Fe_{1-x}Ni_xPS_3$ single crystals were grown using CVT with a temperature gradient from (760-700) to (690-600)°C for 200 hours[89,92]. The magnetic properties of $Fe_{1-x}Ni_xPS_3$ were characterized using temperature and field dependence of magnetization. Figure 6(a) presents the temperature dependence of susceptibility ($\chi = M/H$) with the magnetic field of 1 T along *a*, *b*, and *c**. Here, *c** represents the direction that is perpendicular to the crystallographic *ab*-plane, which is slightly deviated from the crystallographic *c* axis in a monoclinic unit cell. For end compounds $FePS_3$ ($x$ = 0) and $NiPS_3$ ($x$ = 1), susceptibility first increases upon cooling, forming a broad maximum at around temperature $T_{max}$, followed with the AFM transition temperature $T_N$ below $T_{max}$. On increasing Ni content $x$, $T_N$ is found to slightly reduce up to $x = 0.3$, followed by systematic increase with further increasing Ni content [Fig. 6(b)]. Judging from magnetic exchange interaction, such $T_N$ increase can be explained by stronger magnetic exchange interactions $J_1$ and $J_3$ for $NiPS_3$[129] than $FePS_3$[108] (Table 4). Composition dependence for $T_{max}$ follows a similar non-monotonic trend, but $T_{max}$ increases more remarkably with Ni content in Ni-rich compositions [Fig. 6(b)]. As mentioned in the previous section on $Mn_{1-x}Fe_xPS_3$, the broad maximum exists in pristine $MPX_3$ and has been ascribed to the strong low-dimensional dynamic magnetism pertaining to short-range correlated spins. Similarly, in $Fe_{1-x}Ni_xPS_3$ this broad maximum is also suppressed by substitution, leading to sharper susceptibility transition at $T_N$ in substituted samples. This suggests the suppression of the short-range magnetic correlation by substitution in this system. However, the Raman spectroscopy results based on two-magnon Raman scattering reveal that the short-range dynamic magnetism does not quickly quench but rather persists to compositions much below $x$ = 0.9[92]. Therefore, more efforts are needed to clarify the nature of such a broad maximum above $T_N$.

The rapid suppression of a broad maximum with Fe substitution in $NiPS_3$ also suggests the modification of *XY*-type magnetism. In fact, the evolution from *XY*- to Ising-type anisotropy below $x < 0.9$ is clearly manifested in susceptibility. For pristine $NiPS_3$ ($x = 1$), susceptibility measured with field along $a$ ($\chi_a$), $b$ ($\chi_b$), and $c^*$ ($\chi_{c^*}$) axes overlaps above $T_N$ but display significant anisotropy below $T_N$ [Fig. 6(a)], implying a moment orientation along the $a$-axis and consistent with the reported magnetic structure[124]. However, by adding Fe, the $x \leq 0.9$ (i.e., Fe content $\gtrsim$ 10%) samples display anisotropy even above $T_N$, and $\chi_{c^*}$ exhibits much stronger temperature dependence. Those observations are characters for highly anisotropic Ising-type magnetism with out-of-plane moment orientation of the end compound $FePS_3$ ($x = 0$)[108]. Therefore, a transition from *XY*- to Ising-type anisotropy may start with only 10% Fe substitution in $NiPS_3$. Such sensitive tuning of magnetic anisotropy in $NiPS_3$ by just 10% Fe substitution might be attributed to a much stronger ($\approx$ 9 times) single-ion anisotropy ($A$) for $FePS_3$[108] than $NiPS_3$[129] (Table 4).

#### 4.2.1.1.3. $Ni_{1-x}Mn_xPS_3$

$Ni_{1-x}Mn_xPS_3$ provides a platform to study spin-flop transitions under relatively low magnetic fields. In AFM systems, a magnetic field along the easy axis exceeding a critical spin-flop field $H_{SF}$ drives the magnetic moments to rotate to a canted configuration with a component *perpendicular* to the field direction, resulting in a net moment along the easy axis[164,165]. Such field-driven moment reorientation from AFM to SF state is known as the spin-flop (SF) transition, which manifests into a super-linear behavior in field-dependent magnetization measurements (i.e., magnetization upturn). This is different from the spin-flip transition in which the moments are polarized to the magnetic field direction under a strong field and is characterized by a magnetization saturation-like behavior. For $MnPS_3$ with magnetic moment along the out-of-plane direction (Fig. 3)[79,135], a SF transition in the isothermal magnetization has been observed under out-of-plane magnetic field ($H \perp ab$) but is absent under in-plane field ($H//ab$) [Figs. 7(a)-(c)][10,28,65,120,147]. During this SF transition, the magnetic moments undergo continuous rotation towards the *ab*-plane[135]. Although the SF transition was discovered a long time ago[166] and widely studied from bulk to atomically thin $MnPS_3$[29], it has been reported only recently in $NiPS_3$[90]. $NiPS_3$ displays an SF transition under the in-plane field of ~ 6 T at 2 K [Fig. 7(c)], which is consistent with the almost in-plane moment orientation of this compound (Fig. 3)[33,112,167], as shown by a black arrow in Fig. 7(c). Also, the zig-zag AFM structure with moments oriented along *a*-axis leads to an anisotropic SF transition that is sensitive to the in-plane field orientations[90].

The distinct SF behavior in $NiPS_3$ and $MnPS_3$ indicates an evolution of SF transition with composition in "mixed" compounds $Ni_{1-x}Mn_xPS_3$. Indeed, magnetism is found to be highly tunable with composition. Figures 7(b) and (c) show that the in-plane and out-of-plane magnetizations exhibit a systematic enhancement upon increasing Mn content $x$, consistent with the greater magnetic moment of $Mn^{2+}$ than $Ni^{2+}$. Interestingly, the SF transitions are also extremely sensitive to composition variation[90]. The SF transition in $NiPS_3$ under an in-plane field disappears up to 5% Mn substitution (*i.e.*, $x = 0.05$ in $Ni_{1-x}Mn_xPS_3$, Fig. 7(c)). Similarly, the $H_{SF}$ for the out-of-plane field in $MnPS_3$ is reduced by half with 5% Ni substitution (*i.e.*, $x = 0.95$) and disappears upon 10% substitution ($x = 0.9$) [Fig. 7(b)]. Such highly sensitive nature of SF transition to light substitution of magnetic atoms is very different from that of non-magnetic substation. It has been found that 20% substitution is needed to reduce $H_{SF}$ by half in Zn-substituted $MnPS_3$[79,86]. In Ni-substituted $MnPS_3$, the substantially different single-ion anisotropy (Table 4) may explain the efficient suppression of SF transition. A similar mechanism has also been proposed for the suppression of $H_{SF}$ in $MnPS_3$ under pressure[136]. In addition, Ni-Mn substitution can also be viewed as inducing magnetic impurities, especially in low-substitution levels. Hence, it is necessary to consider the magnetic interaction to understand the observed sensitive doping dependence in $Ni_{1-x}Mn_xPS_3$. The magnetic ordering temperature $T_N$ characterizes the strengths of magnetic interactions. For lightly substituted samples with $x$ close to 0 or 1, the $T_N$ is found to only change slightly as compared to the parent compounds $NiPS_3$ and $MnPS_3$, which is interesting given that the light Ni-Mn substitution drastically suppresses the SF transitions. This suggests that the efficient suppression of the SF transition with light magnetic substitution in $Ni_{1-x}Mn_xPS_3$ can be attributed to the tuning of single ion isotropy rather than exchange interaction.

The modification of magnetic anisotropy with Ni-Mn substitution in $Mn_{1-x}Ni_xPS_3$[88] is further demonstrated by the temperature dependence of susceptibility ($\chi = M/H$) [Fig. 7(d)]. For compositions $0 \leq x \leq 0.5$, the out-of-plane susceptibility ($\chi_c$, measured under an out-of-plane field) is significantly reduced below $T_N$ in comparison to in-plane susceptibility ($\chi_{ab}$, measured under an in-plane field), implying moment orientation along the $c$ axis, consistent with the reported magnetic structure of pristine $MnPS_3$[79,135]. This susceptibility trend is reversed on further increasing Ni content above $x > 0.5$ in which $\chi_{ab} < \chi_c$ below $T_N$[88], suggesting switching of magnetic easy axis from out-of-plane to in-plane direction above $x > 0.5$. These results agree well with a theoretical study that predicts the magnetic transition from Néel-type (where all nearest-neighbor spins are aligned antiparallel as shown in Fig. 3) in pristine $MnPS_3$ to zigzag-type in pristine $NiPS_3$ (Fig. 3) at around $x = 0.25$-$0.5$[153].

In addition, the $T_N$ and $T_{max}$ (broad maximum temperature) in susceptibility also provide useful information. As shown in Fig. 7(e), both $T_N$ and $T_{max}$ undergo non-monotonic dependence on Ni-Mn substitution with a minimum around $x = 0.5$[88,90,153] that can be ascribed to the attenuation of the magnetic interaction due to magnetic disorder in such a magnetically substituted system. $T_{max}$ is much larger than $T_N$ for both $MnPS_3$ and $NiPS_3$, but is comparable with increasing substitution levels [Fig. 7(e)], reminiscent to the observation for $Fe_{1-x}Ni_xPS_3$[89,92] described above. Furthermore, for $x = 0.5$ composition, the PM to AFM transition at $T_N = 42$ K is followed by the emergence of a reentrant spin glass state at low temperature[153], which is manifested by the irreversibility between ZFC and FC below $T = 30$ K similar to the case of $Mn_{1-x}Fe_xPS_3$ described above[82].

#### 4.2.1.1.4. $Ni_{1-x}Cr_xPS_3$

The Cr-based layered compounds, such as $CrI_3$, $Cr_2Ge_2Te_6$, $CrCl_3$, and $CrPS_4$, have recently gained intensive attention because of their robust magnetic order persisting in the 2D limit[7,8,168–170]. Surprisingly, the study of Cr-based $MPX_3$ compounds has been very limited. So far, only $CrPSe_3$[171] and $Cr_{2/3}PS_3$[172] have been reported while the stoichiometric $CrPS_3$ has not been experimentally realized. This has been attributed to the weakened P-P dimerization that favors the formation of $Cr^{3+}$ cation instead of the +2 metal valence required to stabilize the $MPX_3$ lattice[173]. The direct substitution of $Cr^{3+}$ for $M^{2+}$ ion in $MPX_3$ is challenging from a chemical valence balance perspective, which may eventually lead to vacancies at metal sites. A similar issue also occurs for $V_xPS_3$ ($x = 0.78$[174,175] and 0.9[71]) in which $V^{2+}$ and $V^{3+}$ ions coexist and lead to V vacancies. Indeed, Cr substitution in $MPX_3$ has been found to be challenging in comparison to other divalent metal ion substitutions[88–90,92,176,177].

So far, light substitution of up to 9% without inducing a clear signature of vacancies in $NiPS_3$ has been recently reported[178]. The lack of vacancies from composition analyses implies an isovalent substitution of +2 valence for Cr ions. Interestingly, a low Cr amount already can efficiently modulate the magnetism in $NiPS_3$. As shown in Fig. 7(f), first, the magnetic transition temperature $T_N$ is found to reduce significantly from 155 K in $NiPS_3$ to 34 K in the $x = 0.09$ sample. Second, the SF transition under an in-plane magnetic field in $NiPS_3$ is significantly suppressed [Fig. 7(f)]. Also, the SF transition starts to occur even under a weak out-of-plane magnetic field of < 1T in the $x = 0.09$ sample, which strongly implies moment reorientation upon Cr substitution. Evaluating the characteristic field for spin flop, magnetic exchange, and magnetic anisotropy also reveals that the magnetic anisotropy is significantly suppressed by Cr substitution. Furthermore, in addition to SF transition, a moment polarization-like behavior has been observed under both in-plane and out-of-plane fields, with a similar polarization field of ~8T and a small saturation moment of 0.24 $\mu_B$ per formula unit.

Suppressed SF transition and more isotropic magnetization are consistent with a suppression of magnetic anisotropy upon Cr substitution, which gives rise to more controllable moment orientations and eventually the moment polarization in the $x = 0.09$ sample. Indeed, similar field-induced moment polarization at relatively low magnetic fields have also been seen in quite a few Cr-based AFM compounds such as $CuCrP_2S_6$[154,155], $CrPS_4$[179], $CrCl_3$[180], and CrSBr[181]. However, owing to the 3+ valence of Cr ions instead of the 2+ metal valence expected in $MPX_3$, substituting Cr is limited to only 9% in $NiPS_3$, as discussed above. This makes it difficult to explore the potential of this route.

Therefore, recently, an alternative approach that balances the valence by co-substituting mono-valent metal ions together with the tri-valent Cr ions has been reported, through which all $M^{2+}$ ions can be replaced to tune the magnetism more efficiently[182]. This work provides a new doping strategy in the $MPX_3$ family, which successfully substitutes an equal Cu and Cr for Ni leading to a wide range of medium-entropy compositions $Cu_xNi_{2(1-x)}Cr_xP_2S_6$ ($0 \leq x \leq 1$).

#### 4.2.1.1.5. $Mn_{1-x}Co_xPS_3$

Compared to other metal substitutions, Co substitution has been fairly elusive in $MPX_3$ with only a few recent studies[115,117,118]. This might be due to the challenges in crystal growth for $CoPS_3$[118] which causes this compound and its derivatives to be less-explored[36,109,130,183]. A recent breakthrough in reactive flux technique using $P_2S_5$ and metal precursors offers an opportunity to study Co-substituted $MPX_3$ compounds[115–119], as described in section 2. The $Mn_xCo_{2-x}P_2S_6$ single crystals can be obtained by annealing the stoichiometric mixture of Mn, Co, and $P_2S_5$ powders at 540°C for 72 hours[118].

The two parent compounds $Mn_2P_2S_6$ (Néel-type AFM with out-of-plane easy axis) and $Co_2P_2S_6$ (zig-zag type AFM with in-plane easy axis) exhibit distinct AFM ordering, so magnetism is expected to be tunable with substituting. As shown in Fig. 8(a), the magnetic susceptibility of $Mn_xCo_{2-x}P_2S_6$ lacks significant irreversibility between ZFC and FC in the entire temperature range, suggesting AFM ordering below $T_N$ without spin-glass state that is also seen in Fe-[82] and Ni-substituted[153] $Mn_2P_2S_6$. Similar to many other metal substituted $MPX_3$[82,86,88,90,153], $T_N$ also exhibits a non-monotonic composition dependence, showing maximum values for the two end compounds ($x$ = 0 and 1) and reduces upon mixing the metal atoms. In addition, the broad maximum temperature $T_{\max}$ observed just above $T_N$ also displays a similar variation with substitution, except that the broad maximum in susceptibility is completely suppressed for the $x = 1$ (50% Mn and 50% Co) sample, as shown in Fig. 8(a). Such suppression is also observed in thin flakes of $Mn_2P_2S_6$, which has been ascribed to the reorientation of $Mn^{2+}$ moments that gives rise to a weak FM state[184].

Similar weak ferromagnetism may also be responsible to the rapid rise in the susceptibility at low temperature below $T < 30$ K [Fig. 8(a)] and non-linear isothermal magnetization at $T = 2$ K [Fig. 8(b)] for this $x = 1$ sample. Nevertheless, one should always be caution about the presence of trace FM impurities in doped materials.

The evolution of Curie-Weiss temperature $\theta_{cw}$ extracted from the linear fit of the high temperature ($T > 200$ K) paramagnetic (PM) region of the inverse susceptibility also provides useful information[118]. $\theta_{cw}$ is negative for AFM $Mn_xCo_{2-x}P_2S_6$, which undergoes a non-monotonic variation with Mn-Co substitution and reaches the maximal (i.e., less negative) at $x = 1$, indicating the attenuation of AFM interaction around this composition. This is probably due to magnetic frustration owing to the competition between distinct AFM exchange interactions $J_1$, $J_2$, and $J_3$ for $Mn_2P_2S_6$[35] and $Co_2P_2S_6$[130].

#### 4.2.1.1.6. $Ni_{1-x}Co_xPS_3$

Magnetic properties of Co-substituted $NiPS_3$ ($Ni_{1-x}Co_xPS_3$) have been investigated in both single crystals and nanosheet (NS) samples[91]. $Ni_{1-x}Co_xPS_3$ single crystals were grown using a CVT technique as shown in Table 1. For NS samples, $Ni_{1-x}Co_xPS_3$ ($0 \leq x < 0.5$) NS of average thickness of 8 nm (around 10 layers) were grown by a chemical vapor conversion method via a two-step process: First, the metal hydroxide NS precursor containing various Ni:Co ratios were synthesized on carbon cloth through a wet chemical method. Second, these metal hydroxide precursors were loaded downstream of an evacuated custom-designed fused silica socket tube with a mixture of P and S positioned on the other side. Ultrathin $Ni_{1-x}Co_xPS_3$ NS can be evenly grown on carbon cloth over a reaction time of 90 min at 480° to 490°C with an argon gas flow [Fig. 8(c)][91].

Magnetization measurements with the background from carbon cloth removed are shown in Fig. 8(d). $T_N \approx 155$ K for pristine $NiPS_3$ is suppressed by Co substitution for both single crystals and NS, which is in line with the general trend of metal substitution in $MPX_3$ as described above. In contrast to bulk single crystals, susceptibility for all $Ni_{1-x}Co_xPS_3$ ($0 \leq x \leq 0.4$) NS samples display low-temperature upturns, suggesting the rise of weak ferromagnetism. Field dependence of magnetization [Fig. 8(d)] also reveals magnetic hysteresis over a low field region ($|H| < 300$ Oe), which becomes more obvious with increasing Co content.

Providing similar lattice parameters and zig-zag type AFM structures for pristine $NiPS_3$ and $CoPS_3$, it is not surprising that Co-Ni substitution does not cause significant modification in magnetic properties. The low-temperature magnetization upturn and magnetic hysteresis occur in NS samples, which might be attributed to the sulfur vacancy in NSs. Electron spin resonance (ESR) spectroscopy measurements[91] have revealed that the sulfur vacancy formation, which is in line with the rise of ferromagnetism with Co substitution. Defect-induced ferromagnetism has been observed in the atomically thin layers of non-magnetic 2D materials such as graphene[2,185] and $MoS_2$[4,186]. As mentioned earlier in section 4.1, in $NiPS_3$, the magnetic interactions are mainly governed by the Ni-S-Ni superexchange interactions, occurring through the overlapping of half-filled Ni $e_g$ and S $3p$ orbitals that mediates the effective electron hopping between neighboring Ni sites[129]. Therefore, tuning of magnetic states due to sulfur vacancy is plausible in $Ni_{1-x}Co_xPS_3$ NS. In fact, the theoretical calculations have demonstrated the disruption of anion-mediated superexchange interaction between Ni moments due to sulfur vacancy[91]. Furthermore, the calculated electronic structure has revealed unoccupied in-gap-like states that could serve as a carrier trap site and hinder the charge transfer process, further suppressing superexchange interaction and might facilitate other competing magnetic phases in $NiPS_3$.

### 4.2.1.1.7. Non-magnetic dopant: $Mn_xZn_{1-x}PS_3$

Magnetic properties of Zn-substituted $MnPS_3$ have been widely studied using magnetization[78,79,86] and Raman[152] measurements. Up on replacing Mn by non-magnetic Zn, the broad maximum immediately above $T_N$ in susceptibility measured with in-plane field starts to vanish with increasing Zn substitution, leading to a sharp susceptibility peak at magnetic transition $T_N$[86]. This suggests a breakdown of short-range correlation, as expected for metal substitution in $MPX_3$[82,88,89,92]. In addition, for Zn-rich samples (Zn content $\geq$ 0.5), the low-temperature susceptibility displays an abrupt enhancement, which has been ascribed to a weak PM behavior of the significant number of isolated spins due to magnetic dilution by Zn[86,152].

Suppression of transition temperature $T_N$ by Zn substitution is seen in both susceptibility[86] and Raman[152] experiments. Unlike magnetic ion substitutions that always lead to a U-shape non-monotonic composition dependency of $T_N$, non-magnetic Zn substitution gradually suppresses $T_N$ to zero temperature up to 50% substitution level in magnetization measurements[78,86] [Fig. 9(a)], or even to 70% from extrapolating the results from Raman spectroscopy[152] [Fig. 9(b)]. This is surprising given the critical concentration for site dilution in a honeycomb lattice with nearest-neighbor interactions ($J_1$) is $x = 0.7$[187]. Therefore, $J_1$ alone may not explain the behavior of $Mn_xZn_{1-x}PS_3$, and the second ($J_2$) and the third ($J_3$) nearest-neighbor interactions, though weaker than $J_1$ in $MnPS_3$ (see Table 4), might be ascribed for the magnetic stability below $x < 0.7$[86]. Of course, measurements on the addition of high-Zn samples would provide a better estimate of the critical composition, which was determined by linear extrapolation of a few compositions[152] as shown in Fig. 9(b).

The impact of Zn substitution in $MnPS_3$ is further manifested as reduced SF transition field $H_{SF}$[79]. Substituting Zn for Mn breaks down the long-range magnetic order, resulting in "weakly bound" Mn moments aligned along the out-of-plane direction. For $MnPS_3$ in which the dipolar anisotropy dominates, the local dipole field along the out-of-plane direction may be attenuated by these weakly bound moments. Furthermore, replacing a Mn with a non-magnetic Zn is accompanied by a vanishing moment at the substitute site, which affects the nearby Mn moments and causes them to be canted[78]. On a larger scale, this leads to an average staggered magnetic moment which consequently suppresses the magnetic anisotropy of the system, which eventually suppresses $H_{SF}$[79].

#### 4.2.1.1.8. Non-magnetic dopant: $Fe_{1-x}Zn_xPS_3$

Magnetic dilution by non-magnetic Zn substitution has also been studied in Ising-type antiferromagnet $FePS_3$[94]. Substituting Zn in $FePS_3$ is reported to induce weak ferromagnetism associated with spin/cluster spin glass[94]. As shown in Fig. 9(c), substituting Fe with non-magnetic Zn suppresses $T_N$. For Zn content $x$ greater than 0.3, significant irreversibility between ZFC and FC measurements starts to occur at temperatures below $T_N$, which has been ascribed to weak ferromagnetism associated with glassy behavior[94]. For high Zn samples ($0.7 \leq x \leq 0.9$), susceptibility upturn appears at low temperatures, which has been attributed to the emergence of FM order with an out-of-plane easy axis[94]. This conclusion is also supported by the Curie-Weiss temperature ($\theta_{CW}$), which becomes less negative with increasing Zn substitution and even attains positive values for $\chi_\perp$ of $x = 0.7$ and 0.9 samples, suggesting the attenuation of AFM interactions and development of FM correlation with Zn substitution in $FePS_3$[94].

The rise of ferromagnetism is further supported by the field dependence of magnetization. The in-plane magnetization displays linear field dependence for all compositions and temperature ranges. However, under $H \perp ab$ magnetic field, the $x = 0.9$ sample displays magnetic hysteresis and saturation behavior with saturation moment reaching ~$4\mu_B$/Fe, which is consistent with the S = 2 $Fe^{2+}$ [94]. The theoretical study further reveals the key role played by hole doping to stabilize ferromagnetism in Zn-substituted $FePS_3$[94]. In highly Zn-substituted samples $x = 0.7$ and 0.9[94], the XPS result reveals the presence of $Fe^{3+}$ ion in addition to $Fe^{2+}$ ion, leading to hole doping and the subsequent rising of FM interactions mediated by bound magnetic polarons induced impurity-band-exchange in such diluted environment induced by Zn substitution in $FePS_3$.

#### 4.2.1.1.9. Non-magnetic dopant: $Fe_{1-x}Cu_xPS_3$

In addition to Zn, magnetic dilution in $FePS_3$ has also been realized by Cu substitution[188]. Cu substitution up to 15% is found to suppress $T_N$ to ≈ 108 K from that of 120 K in pristine $FePS_3$ [Fig. 9(d)][188], consistent with the result obtained for Zn-substituted $FePS_3$[94] and can be attributed to the suppression of magnetic exchange interactions, as discussed in the previous section. Other than that, the susceptibility for the substituted sample is highly similar to that of the pristine $FePS_3$: The out-of-plane susceptibility measured with a magnetic field perpendicular to the basal plane displays a much stronger temperature dependence below $T_N$ than in-plane susceptibility. In addition, in the PM state, susceptibility displays strong anisotropy, with a much higher value for the out-of-plane susceptibility that is similar to undoped $FePS_3$ [Fig. 9(d)]. These observations suggest that magnetic anisotropy is not strongly modified upon Cu substitution up to 15%.

In field-dependent magnetization, the 10% Cu-substituted sample displays a weak non-linearity under an out-of-plane magnetic field [Fig. 9(e)], which was claimed to be a signature of weak ferromagnetism due to the reorientation of canted moments, and has also been observed in Zn-[79] and Ni-substituted[90] $MnPS_3$. In addition to a moment canting scenario, other mechanisms, such as sulfur vacancy which has been seen in Co-substituted $NiPS_3$ nanoflakes[91] (see section 4.2.1.1.5) and the presence of the finite clusters of uncompensated spins[188] might also give rise to weak ferromagnetism in $Cu_{0.15}Fe_{0.85}PS_3$.

#### 4.2.1.1.10. Non-magnetic dopant: $Fe_{1-x}Cd_xPS_3$

Cd as another non-magnetic dopant has been studied a long time ago on a 50% Cd-substituted $Fe_{0.5}Cd_{0.5}PS_3$ sample[189]. Magnetic phase transition is not observed in susceptibility measurements down to 57 K. The Curie-Weiss temperature displays positive values, decreasing from 104 K for pristine $FePS_3$ to 40 K for $Fe_{0.5}Cd_{0.5}PS_3$. The positive Curie-Weiss temperature even in pristine $FePS_3$ appears interesting given it well defined AFM ground state. It might be ascribed to the competition between the direct FM $J_1$ and indirect AFM $J_2$ interactions[189]. The suppression of Curie-Weiss temperature upon introduction of Cd in $FePS_3$ can be attributed to the attenuation of direct $J_1$ owing to the expanded lattice due to Cd substitution, which increases the nearest neighbor distance of $Fe^{2+}$ by 1.95%[189]. A more concrete understanding may be obtained by future substitution studies covering more Cd compositions.

#### 4.2.1.1.11. Metal substitution in selenides: $Fe_{1-x}Mn_xPSe_3$

Most of the metal substitutions were performed on sulfides as summarized above, whereas the selenide $M$PSe$_3$ compounds are relatively less explored. Mn-Fe substitution has been reported in selenide compounds $MnPSe_3$ and $FePSe_3$[87], which are characterized by Néel-type and Ising-type magnetic structures with in-plane and out-of-plane moments, respectively (Fig. 3). Similar to sulfide $FePS_3$ stated above, anisotropy between out-of-plane susceptibility $\chi_\perp$ and in-plane susceptibility $\chi_{//}$ above $T_N$ is a character of Ising-type magnetism in $FePSe_3$. Such anisotropy is suppressed by Mn substitution for Fe [Fig. 10(a)][123], which is expected since the end compounds $FePSe_3$ and $MnPSe_3$ possess different magnetic structures with different easy axes. Interestingly, the susceptibility anisotropy below $T_N$ suggests that the out-of-plane magnetic easy axis for $FePSe_3$ is robust against Mn-substitution up to 90%, which is also supported by the emergence and evolution of the SF transition. SF transition is not observed up to 9T in $FePSe_3$ due to the strong magnetic anisotropy for Ising-type magnetism. Heavy Mn substitution leads the SF to occur in out-of-plane magnetization only, and further increased Mn content beyond 90% causes the SF transition to appear in in-plane magnetization, implying the switching of easy axis from out-of-plane to in-plane when Mn content $x > 0.9$ [Fig. 10(b)]. The rotation of the magnetic easy axis occurring only above 90% Mn for Fe substitution suggests a highly anisotropic magnetic order in $FePSe_3$. This can be estimated from the relative magnitudes of single-ion anisotropy ($A$) for the two end compounds $FePSe_3$ and $MnPSe_3$. Though the experimental value of $A$ is still lacking for $FePSe_3$, in sulfide counterparts $FePS_3$[108] possesses much higher $A$ by > 300 times than that of $MnPS_3$[35] (Table 4). Expecting a similar scenario in selenide samples, the robust magnetic easy axis against up to 90% Mn substitution in $FePSe_3$ can be understood.

Because both end compounds $FePSe_3$[42] and $MnPSe_3$[43] exhibit 2D magnetism down to the monolayer limit, exploring possible 2D magnetism in their mixed compounds ($Fe_{1-x}Mn_xPSe_3$) is interesting. Having said that, the strong frustration arising from mixing two different magnetic metal ions may destabilize magnetic order in the 2D limit. Indeed, such mixing suppresses $T_N$ and leads it to reach a minimum for 50% substitution in studies using both single crystals[123] [Fig. 10(c)] and polycrystals[87] [Fig. 10(d)]. This appears to be a generic behavior for all polymetallic $MPX_3$[82,86,88,90]. In addition, a mixed phase consisting of both $Fe^{2+}$- and $Mn^{2+}$-type ordering, forming nano-sized chemically disordered clusters in intermediate compositions ($0.25 < x < 0.875$) has also been discovered[87] [Fig. 10(d)].

#### 4.2.1.1.12. *HE*P*X*3 (*HE* = High Entropy; *X* = S or Se)

High Entropy (*HE*) alloy involves the mixing of five or more elements each with a composition of 5-35 atomic percentage. The composition of HE compounds can be greatly varied by adjusting the ratios of constituent elements. Such flexible composition stoichiometry allows for property tuning with a wide compositions range. Incorporating multiple elements also offers various valence states and spin states in a system. Depending on the choices of elements, the distribution of charges and spin interactions may vary significantly. Furthermore, the presence of multiple elements results in a complex and disordered atomic structure. Hence, with the added complexity, HE compounds provides an opportunity to investigate the interplay between charge, spin, and composition degrees of freedom. Recently, the bimetallic substitution in $MPX_3$ has been extended to the realm of *HE* compounds[114,122,190]. Providing various anisotropy of ions and diverse magnetic structures for different $MPX_3$ compounds, such a strategy is expected to lead to rich magnetic properties and phases.

For $Mn_{0.25}Fe_{0.25}Co_{0.25}Ni_{0.27}P_{1.04}S_3$[114,190] with equal molar ratio of metal atoms, a well-defined PM to AFM transition at $T_N = 70$ K is found in susceptibility measurements[114] [Fig. 11(a)]. Deviation from this equal metal concentration results in broader or less-obvious AFM transition, as observed in $Mn_{0.2}Fe_{0.3}Co_{0.25}Ni_{0.25}PS_3$[190]. In $Mn_{0.25}Fe_{0.25}Co_{0.25}Ni_{0.27}P_{1.04}S_3$, the magnetic moments are likely to be aligned perpendicular to the basal plan, providing a weaker temperature dependence for susceptibility under an in-plane field. A spin-glass state is also proposed due to the emergence of irreversibility between ZFC and FC susceptibility at a temperature below $T_N$ [Fig. 11(b)]. Similarly, the spin-glass phase has also been observed in a few other *HE*$PS_3$ compounds at different glass temperatures[114,190], as summarized in Table 5. In addition to sulfides, selenide-based HE compounds have also been studied in Fe-Mn-Cd-In-based *HE M*$PSe_3$[122]. The comparison study on $FePSe_3$, $Fe_{0.8}Mn_{0.1}Cd_{0.05}In_{0.05}PSe_3$, $Fe_{0.7}(MnCd)_{0.1}In_{0.1}PSe_3$, and $(FeMnCd)_{0.25}In_{0.17}PSe_3$ has revealed decreased $T_N$ with reducing Fe amount [Fig. 11(c)]. The AFM transition disappears in $(FeMnCd)_{0.25}In_{0.17}PSe_3$, leaving a substantial susceptibility irreversibility around 15 K [Fig. 11(c)] that implies a glass state, which is further confirmed by the frequency-dependent peak in AC susceptibility measurements[122].

#### 4.2.1.2. Non-isovalent substitution

##### 4.2.1.2.1. $CuCrP_2X_6$ ($X$ = S or Se)

The metal substitution in $M_2P_2X_6$ has been extended to the mixing of monovalent $M_1^{1+}$ ($M_1$ = Cu or Ag) and trivalent $M_2^{3+}$ ($M_2$ = V or Cr) ions to from $(M_1^{+1})(M_2^{+3})P_2X_6$[154–158,191–193]. Chemical valence is balanced by maintaining an equal amount of monovalent and trivalent ions. While these compounds have been discovered a long time ago[156,158,191,194], there have been only a few studies on magnetic properties[155,156,158,191,194,195]. $CuCrP_2S_6$[154,155] has recently gained increased attention. In contrast to the random distribution of dissimilar metal ions in isovalent metal-substituted $M_2P_2X_6$[77–92], weak repulsive coulomb interactions and substantial size difference between $Cu^{1+}$ and $Cr^{3+}$ cations[157] favor an alternating arrangements of them in $CuCrP_2S_6$[154,155]: as shown in Fig. 12(a), $Cr^{3+}$ ions are located almost at the center in the vdW layer, whereas Cu+ ions are slightly displaced from the center at the alternative positions above and below the $Cr^{3+}$ layers[155]. This further leads to broken inversion symmetry and ferroelectricity up to room temperature[58,59,196].

Regarding the magnetic properties which is the focus of this review, $CuCrP_2S_6$ displays an AFM order below $T_N$ = 32 K[155][Fig. 12(b)]. On the other hand, $CuCrP_2Se_6$ has a slightly higher $T_N$ (= 40 K) than that of $CuCrP_2S_6$, which is likely attributed to a stronger Cr-Se-Se-Cr superexchange interaction due to enhanced covalency arising from greater atomic orbitals for Se[197]. The much sharper drop in susceptibility measured with the field along the *a*-axis in comparison to other field directions suggests the magnetic moment orientation along the *a*-axis. The linear fit of inverse susceptibility yielded a positive Curie-Weiss temperature $\theta_{cw}$ = 24 K [Fig. 12(b)], implying in-plane FM interaction that may be attributed to an A-type AFM order[155] (see Fig. 3). The effective moments of $\mu_{eff} = (3.78 \pm 0.05)\ \mu_B$/f.u. for $H//ab$ and $\mu_{eff} = (3.89 \pm 0.05)\ \mu_B$/f.u. for $H\perp ab$[154] match with the theoretical value of 3.87$\mu_B$ for $Cr^{3+}$. Like many other Cr-based van der Waals magnets such as $CrPS_4$[179], $CrCl_3$[180], and CrSBr[181], as well as Cr-substitute $NiPS_3$ mentioned above (section 4.2.1.1.6), a field-driven AFM to FM transition has been observed in magnetization measurements with a saturation moment of ≈ 3 $\mu_B$/Cr [Fig. 12(c)]. Consistent with the easy-axis along the *a*-axis, an SF transition appears in magnetization with low $H_{SF}$ ≈ 0.4 T [Fig. 12(d)].

As mentioned above in section 4.2.1.1.4, the mono- and tri-valent of $Cu^+$ and $Cr^{3+}$ allows for co-substitute other bi-valent ions such as $Ni^{2+}$ to form $Cu_xNi_{2(1-x)}Cr_xP_2S_6$ ($0 \leq x \leq 1$)[182]. This doping strategy provides further tuning of magnetism in the $MPX_3$ family and leads to a wide range of medium-entropy compositions.

#### 4.2.1.2.2. Ag$MP_2X_6$ ($M$ = V or Cr; $X$ = S or Se)

$AgVP_2S_6$ and $AgCrP_2S_6$ have been studied a long time ago[192,193]. Structurally, these materials form a zig-zag chain of $V^{3+}$ ($S = 1$) and $Cr^{3+}$ ($S = 3/2$) ions respectively along the $a$-axis [Fig. 13(a)], which is distinct from other $MPX_3$ compounds. Long-range AFM order is absent in $AgVP_2S_6$ from 5 to 200K in susceptibility measurements[192] [Fig. 13(b)]. In contrast, $AgCrP_2S_6$ is reported to be AFM below $T_N \approx 7$ K[193] [Fig. 13(c)]. Such $T_N$ from susceptibility is lower than $T_N \approx 20$ K obtained from powder neutron diffraction experiment[156]. Both compositions display enhanced susceptibility at low temperatures, but $AgVP_2S_6$[192] is characterized by isotropic susceptibility for magnetic field directions along and perpendicular to the zig-zag chain direction [Fig. 13(a)], while $AgCrP_2S_6$ features a small anisotropy at low temperatures[193] [Fig. 13(b)].

A more recent study has expanded to selenide counterpart $AgVP_2Se_6$[198] and $AgCrP_2Se_6$[199]. These compounds are structurally slightly different from the above $AgVP_2S_6$ and $AgCrP_2S_6$ [Fig. 13(d)]. Interestingly, susceptibility irreversibility and magnetization saturation of $AgVP_2S_6$[198] [Fig. 13(e)] implies ferromagnetism with a $T_c \sim 18.5$ K, which is distinct from the AFM ground state for other $MPX_3$. On the other hand, $AgCrP_2Se_6$[199] exhibits AFM ground state below $T_N \sim 42$ K, which features field-driven moment polarization under the high field of $B \sim 3$-$6$ T. Furthermore, polar-reflective magnetic circular dichroism (RMCD) experiments on $AgVP_2Se_6$[198] found a growing hysteresis loop with reducing thickness in thin flakes, which persist down to 6.7 nm flake thickness with unchanged $T_c$.

### 4.2.2. Chalcogen ($X$) substitution

As discussed in section 3, transition metal atoms $M$ are surrounded by $P_2X_6$ clusters in each layer of $MPX_3$ (Fig. 2). So, substituting chalcogen $X$ modifies the local environment of $M^{2+}$ ion within honeycomb layers with the magnetic plane intact [Fig. 4(b)]. Therefore, chalcogen substitution offers a relatively clean approach to modifying magnetic exchange interactions. This approach has been demonstrated to be effective in other vdW magnets such as chromium halides[146,147], in which the magnetism can be efficiently controlled by non-magnetic ligand substitution. Although the syntheses and structural characterizations of chalcogen-substituted $MPX_3$ have been reported a long time ago[200–202], their magnetic properties were studied recently in a few Mn-, Fe-, and Ni-based $MPX_3$ compounds[31,32], as described below.

#### 4.2.2.1. $MnPS_{3-x}Se_x$

In $MnPS_{3-x}Se_x$, the successful substitution can be readily seen by the colors of the crystal. Upon increasing the Se content, these relatively transparent crystals gradually change color from green to wine red [Fig. 14(a)], indicating the variation of the optical gap[31]. Upon Se substitution, $T_N$ displays a monotonic, slight reduction from 78.5 K for $MnPS_3$ to 74 K for $MnPSe_3$[31,32] [Fig. 14(b)], which is drastically different from non-monotonic composition dependence seen in metal-substituted $MnPS_3$[82,86,88,90]. This is found to be caused by the suppressed exchange interaction[31]. Magnetism in $MnPS_3$ is governed by the nearest-neighbor exchange interaction ($J_1$) from the Mn-Mn direct exchange[39,129,130], and the third nearest-neighbor interaction ($J_3$) is also considerable (Table 5). Se substitution for S expands in-plane lattice[203,204] and elongates Mn-Mn distance, which attenuates the direct exchange and consequently reduces $T_N$[31,32].

In addition to magnetic exchange interactions, chalcogen S-Se substitution also modifies magnetic anisotropy[31]. This is expected because the two end compounds $MnPS_3$ and $MnPSe_3$, though both possess Néel-type AFM structure within each vdW layer where all nearest-neighbor spins are aligned antiparallel, feature different magnetic out-of-plane[79,135] and almost-in-plane[30,87,95,96] easy axis, respectively[30,79,87,95,96,135] (Fig. 3). The modification of magnetic anisotropy is reflected by the evolution of SF transition. SF transition in $MnPS_3$ occurs under an in-plane field. The SF field $H_{SF}$ is gradually suppressed to zero upon increasing Se content near $x = 1$ ($MnPSSe_2$). On the other hand, the SF transition re-appears but under an out-of-plane field for $x > 1$ samples. Since SF transition in AFM material is driven by magnetic field component along the magnetic moment direction (i.e., easy axis), this implies the rotation of easy axis from out-of-plane towards the in-plane direction with Se substitution, which can be attributed to the enhanced single-ion anisotropy upon Se substitution that usually favors the in-plane moment orientation in $MPX_3$. The switching of the easy axis is found to happen closer to the $MnPS_3$ side i.e., between $x = 0.7$ and 1.2 [Fig. 14(c)], implying the magnetism in $MnPS_3$ is softer than $MnPSe_3$, which is also consistent with the results obtained from the theoretical calculations that reveal much smaller anisotropy between the in-plane and out-of-plane directions of the spins in $MnPS_3$ than $MnPSe_3$[31].

#### 4.2.2.2. $NiPS_{3-x}Se_x$

The study on Ni-based $NiPS_{3-x}Se_x$ faces difficulty in synthesizing single crystals for Se-rich compositions. Indeed, the studies on $NiPSe_3$ are very limited[39] as compared to the sulfide counterpart $NiPS_3$. A recent investigation on $NiPS_{3-x}Se_x$[31] combining single crystal with Se content up to $x = 1.3$ [Fig. 14(d)] and polycrystals up to $x = 3$ have revealed monotonic enhancement in $T_N$ with increasing Se content, from ~155 K for $NiPS_3$ to ~210 K for $NiPSe_3$ [Fig. 14(e)]. This observation is distinct from the reduced $T_N$ due to metal substitution in $NiPS_3$[88,90]. Interestingly, compared to $MnPS_{3-x}Se_x$ in which Se substitution slightly reduces $T_N$[31,32], in $NiPS_3$ the similar Se substitution causes strong $T_N$ enhancement[31]. Such results have been ascribed to different governing exchanges in Mn- and Ni-based systems[31]. Magnetism of both systems are determined by the nearest $J_1$ and third nearest $J_3$ while the second nearest-neighbor interaction $J_2$ is negligible (Table 4). Unlike $MnPS_3$ in which the direct exchange between the neighboring metal ions contributes to $J$ as described in the previous section, neutron scattering experiments[39,129] have demonstrated that the magnetic interactions in $NiPS_3$ are superexchange in nature. The direct exchange between $Ni^{2+}$, however, does not exist due to the filled $t_{2g}$ orbitals for $Ni^{2+}$[39,129]. Therefore, compared to S, Se with extended $4p$-orbitals favors orbital overlap and enhances superexchange between $Ni^{2+}$ through chalcogen, giving rise to enhanced $T_N$.

Similar to the case of $MnPS_3$, Se substitution also tunes SF transitions in $NiPS_3$[31]. The SF transition under an in-plane field of ~ 6 T for $NiPS_3$ ($x$ = 0) occurs at a higher in-plane field of ~8 T for $x$ = 0.25 [Fig. 14(f)]. Further increasing Se content up to 1.3 results in essentially linear magnetization under both in-plane ($H//ab$) and out-of-plane ($H\perp ab$) fields up to 9 T. This is consistent with enhanced exchange interactions with Se substitution, based on evaluating the SF field, $H_{SF} \approx \sqrt{2H_E H_A}$ where $H_E$ and $H_A$ are effective exchange and magnetic anisotropy fields, respectively[205]. SF transition also provides a measure of magnetic anisotropy. When SF transition switches from one magnetic field direction to another, it suggests rotation of magnetic easy axis as seen in aforementioned $MnPS_{3-x}Se_x$. For $NiPS_{3-x}Se_x$, however, unavailable single crystals for Se-rich compositions, and the lack of experimentally determined magnetic structure for end compound $NiPSe_3$, make it difficult to judge the evolution of magnetic anisotropy based on magnetization study.

#### 4.2.2.3. $FeP(S_{1-x}Se_x)_3$

In the $MPX_3$ family, $FePS_3$ has been identified as a representative material that displays Ising-type magnetism characterized by out-of-plane magnetic moments (Fig. 3)[6,40,108]. Such Ising-type magnetism in $FePS_3$ has been proposed to originate from the combination of the strong spin-orbit coupling (SOC) of the high-spin $Fe^{2+}$ ($d^6$) state and the trigonal distortion of $FeS_6$ octahedra[40]. Another Fe-based $MPX_3$ compound $FePSe_3$ which was discovered a few decades ago[30,101] has so far received surprisingly less attention than $FePS_3$ despite the fact that both $FePS_3$[6,40,108] and $FePSe_3$[30,42] exhibit similar Ising-type zig-zag AFM ordering from bulk to the monolayer limit. This is in stark contrast to the distinct magnetic structures (experimentally or theoretically determined) for sulfide and selenide compounds in some other known $MPX_3$ such as $MnP(S,Se)_3$ and $NiP(S,Se)_3$[31,206]. Such lack of change in magnetic structure in $FeP(S,Se)_3$ naturally raises a question of whether chalcogen substitution may play any role in modifying magnetic properties, which has been investigated in Ref. [123]. For the entire composition range, magnetic susceptibility displays strong anisotropy both below and above $T_N$, with the in-plane susceptibility measured under the in-plane field ($H//ab$) showing much weaker temperature dependence as compared to the out-of-plane susceptibility [Fig. 15(a)], implying the persistence of the Ising-type magnetism up on Se substitution. In Mn- and Ni- based $MPX_3$, the magnetic anisotropy mainly originate from anisotropic superexchange interaction that arises from SOC of non-magnetic ligands[207,208]. However, in $FeP(Se_{1-x}S_x)_3$, magnetic anisotropy predominantly stems from the strong crystal electric field on $Fe^{2+}$ ion (discussed in section 4.1.)[139,140], and thus S-Se substitution does not substantially change magnetic anisotropy and consequently maintains the robust Ising-type magnetic ordering in $FeP(Se_{1-x}S_x)_3$[123].

The evolution of $T_N$ with Se substitution is also interesting. Unlike Mn- and Ni-based $MPX_3$ in which chalcogen substitution leads to monotonic decrease or increase in $T_N$, respectively (see sections 4.2.2.1 and 4.2.2.2), in Fe-based system, a non-monotonic dependent is observed[123] [Fig. 15(b)]. This has been attributed to the evolution of magnetic exchange[123]: Previous neutron scattering measurement has revealed dominant $J_1$ in $FePS_3$[108] (Table 4), which is FM in nature and sensitive to Fe-Fe distance due to direct exchange interaction of the nearest-neighbor Fe moments[129,189]. A similar direct exchange scenario occurs for $MnPS_{3-x}Se_x$ except $J_1$ is AFM in nature (see sections 4.2.2.1). With the consideration that the competition between FM and AFM correlations determines the magnetic ordering type and transition temperature, the evolution of $T_N$ can be understood in terms of the nearest-neighbor Fe-Fe distance that governs the FM $J_1$ in $FeP(Se_{1-x}S_x)_3$. As shown in Fig. 15(c), the nearest-neighbor Fe-Fe distance in $FeP(Se_{1-x}S_x)_3$ first reduces with S substitution up to $x = 0.5$. Further increasing S content slightly modifies the lattice structure and leads to elongated Fe-Fe distance. This causes a non-monotonic composition dependence for $J_1$ and eventually results in non-monotonic evolution for $T_N$.

#### 4.2.3. Inter-layer intercalation

In addition to metal and chalcogen substitutions, the layered structure of $MPX_3$ allows for another doping strategy - inter-layer intercalation of guest ions, as shown in the schematic in Fig. 4(c). The metal and chalcogen substitutions are all based on the direct mixing of dopants and source materials during synthesis such as flux and chemical vapor transport (CVT). On the other side, the intercalation technique is usually applied to post-grown crystals, involving the insertion of guest ions into inter-layer spaces of the material. So, this method is relatively clean and does not strongly modify the host layers. Generally, intercalation to vdW materials is expected to modulate the inter-layer spacing and may also lead to carrier doping. Therefore, compared to metal $M$ and chalcogen $X$ substitutions, intercalation offers a new approach to tuning the magnetic properties of $MPX_3$ materials, which may create new magnetic phenomena and novel magnetic phases in $MPX_3$ that are not accessible by substitution. Intercalation of Li[54,99,101,209,210] and various organic-ions[97,98,100,102] have been reported in $MPX_3$, which were achieved by electrochemical[54,98–100,209,210] and wet chemical[97,101,102] techniques.

#### 4.2.3.1. Lithium intercalation

Lithium (Li) intercalation in $MPX_3$ has been reported a long time ago[54,101,209,210]. Earlier work mainly focused on the electrical, optical, and electrochemical properties of Li-intercalated $MPX_3$. Regarding magnetic properties, the effects of Li intercalation on $NiPS_3$, $FePS_3$, $FePSe_3$, and $MnPSe_3$ compounds were first studied four decades ago[101]. The intercalation was achieved by reacting $MPX_3$-powered samples and *n*-butyllithium solution in hexane at room temperature in a dry, oxygen-free environment for 10 – 15 days. The duration of the reaction and concentration of *n*-butyllithium was found to affect the amount of intercalated Li. In this work[101], resistivity reduction in those powered samples implies successful intercalation. However, magnetic properties including transition temperature, effective moment, and Weiss temperature extracted from temperature-dependent magnetic susceptibility measurements do not display notable changes [Fig. 16(a)].

Recent studies adopting electrochemical interaction to intercalate Li in single crystalline samples reveal modified magnetic properties in $Li_xNiPS_3$[99] and $Li_xFePS_3$[211]. The intercalation in both studies was performed using a battery setup [Fig. 16(b)] and the expanded *c*-axis implies successful intercalation. For $NiPS_3$, samples with Li content up to $x = 0.4$ are air-stable and maintain good crystallinity based on x-ray diffraction and scanning electron microscope imaging, though non-magnetic $Li_3PS_4$ impurity starts to occur for $x \geqslant 0.3$. It is worth noting that the Li amount is the nominal content calculated based on the intercalation duration and electrical current because neither x-ray diffraction nor energy dispersive x-ray spectroscopy can directly probe Li.

Similar to the earlier study using *n*-butyllithium solution for Li intercalation[101], the electrochemically intercalated $Li_xNiPS_3$ ($x$ = 0 - 0.4) samples do not display clear changes in $T_N$. However, the development of non-linear field dependence for magnetization, and its saturation-like behavior after removing the linear background [Fig. 16(c)], and the rise of the hysteresis loop and tiny magnetic moment [Fig. 16(d)] have been ascribed to ferrimagnetism due to uncompensated antiparallel magnetic moments[99]. For $FePS_3$[211], in contrast, $T_N$ is found to be reduced with the expansion of the c-axis by Li-intercalation, suggesting a possible role of inter-layer magnetic exchange. The Ising-type magnetism characterized by strong anisotropy in magnetic susceptibility in the high-temperature paramagnetic state remains robust. At low temperatures, non-linear field dependence for magnetization is also observed, similar to Li-intercalated $NiPS_3$.

#### 4.2.3.2. Organic-ion intercalation

The vdW gap allows for not only the intercalation of small Li atoms or ions but also the much larger organic molecules or ions[97,98,100,212,213]. Organic ion intercalation has been mainly performed on $NiPS_3$[97,98,100]. Intercalating tetraheptyl ammonium ($THA^+$)[98], tetrabutylammonium ($TBA^+$)[100], cobaltocenium ions [$Co(Cp)_2^+$ , where Cp is a cyclopentadienyl ring $C_5H_5^-$][100], and 1,10-phenanthroline[97] using electrochemical[98,100] and solution reaction[97] have been performed. Compared to Li-intercalation, intercalating large organic ions results in a much more significant elongation of inter-layer spacing[97,98,100]. It has also been found that the intralayer lattice structure does not display remarkable change upon intercalation[97].

Organic-ions intercalation in $NiPS_3$ is found to drastically modify the magnetic properties[97,98,100], mainly giving rise to a low-temperature FIM state[98,100] that is also reported in Li-intercalated $NiPS_3$[99] described in the previous section. The conclusion has been drawn based on the transition-like sharp susceptibility upturn around and below 100 K [Figs. 17(a) and 17(c)], as well as magnetic hysteresis loops in $THA^+$-[98], $TBA^+$-[100], and $Co(Cp)_2^+$[100]-intercalated $NiPS_3$ [Figs. 17(b) and 17(d)]. It has been proposed[100] that the cation intercalation causes the reduction of certain Ni from the $Ni^{2+}$ to the $Ni^0$ state, together with its displacement from the $O_h$ to the $T_d$ lattice site. Such transition from high spin $Ni^{2+}$-$O_h$ atoms ([Ar]$3d^8s^0$) to zero-spin $Ni^0$-$T_d$ ([Ar]$3d^{10}s^0$) leads to uncompensated antiferromagnetism, i.e. a FIM state. An alternative mechanism based on the Stoner effect has also been proposed[98]: The charge doping due to organic-ion intercalation enhances the density of states (DOS) at the Fermi level and eventually triggers the inter-chain Stoner splitting of the itinerant electrons, resulting in the higher electron concentration in one Ni chain to create net magnetic moment which gives rise to ferrimagnetism.

Recently, a new method using iron dopant as reaction active sites to intercalating organic-ion in $NiPS_3$ has been reported[97]. In this method, light Fe-substituted $NiPS_3$ (i.e., $Fe_{0.02}Ni_{0.98}PS_3$) single crystals are synthesized, which is followed by intercalation of a complexing agent 1,10-phenanthroline by solution reaction with aniline chloride[97]. In this process, 1,10-phenthroline removes $Fe^{2+}$ ions into the solution, creating metal-ion vacancies. The aniline chloride provides a proton to 1,10-phenthroline, and such protonated 1,10-phenanthroline is attracted towards the vacant metal sites, forcing intercalation of 1,10-phenanthroline into inter-layer spacing. The intercalated samples also display the signature of ferrimagnetism.

## 5. Engineering magnetism through pressure

In addition to chemical doping, pressure has also been established as an efficient external stimulus to tune magnetic properties. In $MPX_3$, high pressure has been found to effectively modulate electronic properties[64–72]. Recently, pressure has also been used to control magnetism in this material family[214–217]. A recent review article has extensively summarized the theoretical and experimental progress of the structural, electronic, magnetic, and optical properties tuning of $MPX_3$ under pressure[218]. In this review, we focus on engineering magnetism in $MPX_3$ materials. Since pressure generally tunes lattice structures and possibly induces structure transitions, it also sheds light on tuning magnetism through changing magnetic exchange and anisotropy. Therefore, here we briefly describe the pressure-driven modification of magnetism in $MPX_3$. The magnetic properties of $MPX_3$ are found to be sensitive to high pressure. $T_N$ systematically increases with the pressure for $MnPS_3$[214,217] and $FePS_3$[215], which can be explained by the enhanced magnetic interactions under pressure. In particular, compressing the crystal planes closer strengthens the inter-layer interaction[215,217]. For $MnPS_3$, the $T_N$ also enhances with applying pressure, with different reported rates of 6.7 K/GPa[214] and 13 K/GPa[217] depending on the type of samples (polycrystal or single crystal). In addition, SF transition is tunable by pressure. In $MnPS_3$[217], the SF field decreases at a rate of $4.0 \times 10^3$ Oe/GPa, which is likely due to the reduction of magnetic anisotropy under pressure. Furthermore, the magnetic structure for $MnPS_3$ remains unchanged up to ~3.6 GPa[214] but changes at ~2 GPa for $FePS_3$[215] and ~4 GPa for $NiPSe_3$[216]. Such changes in magnetic structures are accompanied by the insulator-to-metal transition and even superconductivity in of $FePSe_3$[72] and $NiPSe_3$[216]. Therefore, high pressure, as well as strain, are excellent tools that can be combined with chemical tuning (substitution and intercalation) to achieve additional properties, such as the combination of the desired magnetic structure and electrical conductivity.

## 6. Summary and perspectives

In this review article, we provide a summary covering the crystal growth, structure, and magnetic properties of the vdW-type AFM $MPX_3$ material family, with a focus on property tuning by various doping scenarios. Metal substitution, chalcogen substitution, and inter-layer intercalation are highly effective in tuning the magnetic properties in $MPX_3$, including magnetic ordering, exchange interactions, magnetic anisotropy, etc. Generally, AFM ground states are robust, and most of those metal and chalcogen substitution tune properties within antiferromagnetism, except for a few cases such as $Fe_{1-x}Zn_xPS_3$ and $AgVP_2Se_6$ that display FM-like behaviors such as irreversibility, hysteresis, and saturation magnetization. However, magnetization is not a bulk measurement as a tiny number of magnetic impurities may give rise to a strong magnetization signal. Also, quite a few FM-like signatures may originate from a spin-flip transition or a canted AFM state, as seen in AFM $CuCrP_2S_6$ and $Ni_{1-x}Cr_xPS_3$. Therefore, other bulk measurements such as neutron diffraction are needed to clarify the magnetic ground state as well as any field-induced phase transitions. Indeed, the rapid development of $MPX_3$ and its derivatives calls for more neutron resource allocations on these interesting vdW magnetic compounds.

On the other hand, intercalations of a few guest species have been reported to result in a FIM state at low temperatures. The differences in substitution and intercalation might be attributed to different consequences: substitution within each individual layer plays a role mainly in altering magnetic anisotropy and exchange interactions while intercalation in the inter-layer spacing might lead to carrier doping or reductive reaction for metal ions. Also, intercalation is efficient in tuning inter-layer distance, especially when intercalating large ions such as organic intercalation. This effectively reduces interlayer interaction and may push the magnetism approach to a 2D limit.

Therefore, intercalation might be a good tool to engineering magnetism for compounds in which the interlayer exchange $J_z$ is important. So far, there are not many reported intercalation studies, therefore more efforts toward this direction are needed to fully explore this promising material tuning approach. Similarly, a direct probe for magnetic ordering such as neutron diffraction is also highly desired in this direction. Another urgent need is an accurate and convenient approach to determine the amount of Li for Li-intercalated samples.

High-entropy (*HE*) $MPX_3$ is an emerging field with great opportunities owing to the expanded degrees of freedom. The combination of multiple elements in *HE* compounds offers greater versatility and tunability of material properties as compared to conventional bi-metallic and bi-chalcogenide alloys by varying the composition of *M* and *X* elements. Incorporating multiple components in an *HE* compound could induce a severe lattice distortion due to the difference in their atomic sizes, which provides opportunities for crystal structure and symmetry engineering for possible novel functional properties. The *HE* alloys are expected to generate unusual magnetic orderings due to competing magnetic interactions enabled by the interplay between spin, charge, and orbital degrees of freedom[219]. On the other side, the randomness introduced by high entropy may also quench the magnetic order and simultaneously facilitate the insulator-to-metal transition and subsequent emergence of superconductivity, which may be extremely challenging but has been observed in $FePSe_3$-based *HE* compounds under external pressure[122]. Further exploring *HE* $MPX_3$ compounds could provide pathways to realize enhanced conductivity and even superconductivity at an ambient pressure or much lower pressure. Understanding the effect of entropy on the structural stability and material properties of complex *HE* compounds is critical for their optimal design and utilization in practical applications. However, the controlled synthesis of the alloy composition, and the large parameter space, pose significant challenges for controlled study (e.g., fixing the content of one or more metals) to clarify the mechanism behind the property changes. More effort, especially in synthesis and applying additional experimental tools, would be beneficial.

One driving force of the study of $MPX_3$ is the feasibility of obtaining their atomically thin layers that provide invaluable opportunities for studying new phenomena originating from 2D magnetism and future device applications. The tunable magnetism summarized in this review further demonstrates the great potential of this material platform. Despite their well-established magnetic orders in bulk materials and the fact of being one of a few first demonstrated 2D magnets in the atomically thin layer form, the study of 2D magnetism in $MPX_3$ is still in an early stage with only a few pristine compositions being investigated, as summarized in Table 6. It has been found that the long-range magnetic orders in atomically mono- or bi-layers only exist in a few $MPX_3$ compounds such as $MnPS_3$[29,110], $FePS_3$[6], $NiPS_3$[60], $CoPS_3$[109], $MnPSe_3$[43], and $FePSe_3$[42]. Whereas the 2D magnetism in substituted and intercalated $MPX_3$ compounds remains elusive. So far, few challenges to exploring magnetism in 2D have been identified in the $MPX_3$ family. First, because of the absence of net magnetization, direct probe of AFM states in 2D layers of $MPX_3$ is challenging by using nanoscale magnetic characterization techniques such as scanning single-spin magnetometry[220], magneto-optical Kerr effect microscopy[7,8], polar reflective magnetic circular dichroism (RMCD)[10], x-ray magnetic dichroism (XMCD)[168], and optical linear dichroism (LD)[48,221]. In addition to developing more experimental techniques for probing AFM state in nanoscale, inducing ferromagnetism in $MPX_3$ can be a shortcut for 2D magnetism study. We hope the various doping techniques summarized in this review can provide some insight into creating FM ground states for field-driven polarized FM states in these materials. Another advantage of inducing ferromagnetism is the breaking of time-reversal symmetry that may bring in other exotic phenomena. Recent studies have predicted the non-trivial topological states in the FM phase of $MPX_3$[222,223], providing a rare platform for investigating the interplay between magnetism and band topology in the 2D limit.

Although this review mainly focuses on magnetism, we wish to take this opportunity to comment on the electronic properties of $MPX_3$. Those materials are well-known wide band gap insulators and electronic transport measurements in bulk materials are almost impossible or only possible in the high-temperature range[71,198]. Enabling electronic studies in vdW-type magnetic $MPX_3$ could be an important direction. Exploiting the coupling between magnetic and electronic properties would provide a platform for tunable electrical properties to develop quantum materials and devices with multifunctional magnetic and electronic properties. Therefore, though applying high pressure has been demonstrated to be effective[64–72], further development in approaches or materials that are compatible with 2D magnetism and device integration is critical, which calls for synergistic efforts in material modeling, high-throughput calculations, synthesis, characterization, and device study.

The vdW structure of $MPX_3$ materials is naturally advantageous for fabricating devices for electronic and optoelectronic applications. However, the highly insulating nature of $MPX_3$ materials has somewhat hindered the rapid development of device applications. To the best of our knowledge, so far the field-effect transistors (FET) using atomically thin layers of only a few $MPX_3$ compounds such as $NiPS_3$[224] and $MnPS_3$[53] have been reported. The FET based on $NiPS_3$[224] and $MnPS_3$[53] thin flakes have been found to exhibit $n$-type and $p$-type semiconducting behavior with on/off ratios of ~ $10^3$-$10^5$ and ~ $10^2$-$10^3$ respectively at room temperature, with low charge carrier mobilities of ~ 0.5-1 $cm^2V^{-1}s^{-1}$ and $8.3 \times 10^{-3}$ $cm^2V^{-1}s^{-1}$ respectively. On the other hand, recent works have mainly focused on optoelectronic devices such as photodetectors[52,225–227], phototransistors[228,229], and sensors[230] based on a few layers of $MPX_3$. High-performance and sensitive photodetectors working in a wide frequency range from near-infrared[227] to ultraviolet[52] have been successfully fabricated using the thin layers of $MnPS_3$[53], $FePS_3$[52,225], $MnPSe_3$[226], and $FePSe_3$[227]. The few layers $NiPS_3$[229] and $MnPSe_3$[228] have also been integrated into phototransistors for ultrasensitive light detection. Furthermore, the $MnPS_3$ thin flakes[230] has been demonstrated as sensors for sensitive and selective moisture sensing. Providing this progress, efficiently tuning the magnetism would possibly lead to new optoelectronic and photonic applications, especially the light polarization-based applications for which magnetism can play a role. Furthermore, metalizing $MPX_3$ via combined chemical and strain may pave a way for electronic applications.

Finally, more attention should be given to atomic ordering in these substituted and intercalated compounds. The distribution of the dopants and possibly the associated vacancies may play important roles in mediating properties. The ultimate examples are well-defined atomic orderings such as superlattice or Peierls transition, which can be relatively easily probed by high-resolution structure characterization tools such as High-resolution/ Scanning transmission electron microscopy (HRTEM/STEM) and synchrotron XRD. However, chemical short-range order, which is another ultimate example, is difficult to characterize and has not been widely studied in crystalline solids. Such chemical short-range order is distinct from the well-known short-range order in condensed matter physics that is defined from the long-range ordering. Roughly speaking, the long-range order is based on periodic ordering with long-range correlation length, while the short-range order is normally referred to as similar ordering but with much-shortened correlation length, which usually occurs with the disturbance of the long-range order by doping or increasing the temperature to around the phase transition temperature. In contrast, chemical short-range order occurs in a doped system and characterizes the deviation of the atomic distribution from the *perfect random distributions*. The study started from the III-V semiconductor alloys decades ago which mainly focused on long-range order[231], and has been recently extended to chemical short-range order in other systems such as group-IV semiconductors[232,233], metal alloys[234,235], thermoelectric materials[236,237], and oxides[238]. Thus far, chemical short-range order remains to be an extremely less-explored topic in condensed matter physics, probably owing to the lack of efficient experimental tools to probe it. In doped $MPX_3$ systems, chemical short-range order should also occur and may affect the magnetic properties. Providing rich and tunable magnetism, the doped $MPX_3$ compounds provide an ideal platform for studying the nature of the possible various chemical short-range orders in vdW materials and their impacts on magnetism, which is a new field awaiting exploration.

**Author contributions**

R. B. and J. H. designed, wrote, and edited this manuscript.

**Conflicts of interest**

The authors declare no conflict of interest.

**Acknowledgement**

R. B. acknowledges the support from μ-ATOMS, an Energy Frontier Research Center funded by DOE, Office of Science, Basic Energy Sciences, under Award DE-SC0023412. J. H. acknowledges the support from the U.S. Department of Energy, Office of Science, Basic Energy Sciences program under Grant No. DE-SC0022006. The authors are grateful to T. Li at GWU for informative discussions.

## Tables

**Table 1:** Summary of crystal growth of $MPX_3$ compounds.

| Materials | Growth method | Temperature (°C) | | Duration (days) | Ref. |
|---|---|---|---|---|---|
| | | $T_{hot}$ | $T_{cold}$ | | |
| $VPS_3$ | CVT | 600 | 500-350 | 7-30 | 71,107 |
| $MnPS_3$ | CVT | 780-650 | 720-600 | 7 | 31,41,88,105,239 |
| | $P_2S_5$ Flux | 650 | - | 2 | 116 |
| $FePS_3$ | CVT | 750-700 | 730-600 | 9-21 | 6,72,89,92,105,108,240 |
| | $P_2S_5$ Flux | 650 | - | 3 | 115–117 |
| $CoPS_3$ | CVT | 600 | 550-500 | 7-8 | 109,241 |
| | $P_2S_5$ Flux | 580 | - | 2 | 115–117 |
| $NiPS_3$ | CVT | 760-670 | 690-550 | 7-21 | 31,89,90,92,124 |
| | $P_2S_5$ Flux | 650 | - | 2 | 116 |
| $ZnPS_3$ | CVT | 600 | 550 | 7 | 152 |
| | $P_2S_5$ Flux | 650 | - | 2 | 116 |
| $CdPS_3$ | CVT | 630 | 600 | 5 | 105 |
| | $P_2S_5$ Flux | 650 | - | 2 | 116 |
| $MgPS_3$ | CVT | 670 | 550-800 | 90 | 125 |
| | $P_2S_5$ Flux | 650 | - | 2 | 116 |
| $PdPS_3$ | | 500 | 450-500 | 7 | 125 |
| $SnPS_3$ | | 630 | 600 | 3-5 | 105,106 |
| $HgPS_3$ | CVT | 350 | 320-500 | 3 | 125 |
| | $P_2S_5$ Flux | 400 | - | 2 | 119 |
| $PbPS_3$ | | 650 | 620 | 3-5 | 106 |
| $CrPSe_3$ | | 800 | 700 | 10 | 104 |
| $MnPSe_3$ | | 650 | 600 | 7 | 31,242 |
| $FePSe_3$ | | 720-700 | 700-600 | 7-10 | 72,243 |
| $NiPSe_3$ | | 530 | 500-560 | 90 | 125 |
| $ZnPSe_3$ | | 370 | - | 10 | 244 |
| $CdPSe_3$ | | 610 | 650-550 | 30 | 125 |
| $MgPSe_3$ | | 620 | 480-690 | 90 | 125 |
| $HgPSe_3$ | | 400 | 300 | 5 | 245 |
| $PbPSe_3$ | | 750 | 710 | 3-5 | 106 |
| $CuCrP_2S_6$ | | 750-720 | 700-680 | 5-8 | 154,155 |
| $CuInP_2S_6$ | CVT | 800 | 700 | 7 | 246 |
| | $P_2S_5$ Flux | 650 | - | 4 | 116 |
| $AgCrP_2S_6$ | | 750 | 690 | 5 | 157 |
| $AgInP_2S_6$ | | 750-680 | 610-600 | 7-9 | 247,248 |
| $Mn_{1-x}Fe_xPS_3$ | | 730 | 700-630 | 5 | 82 |
| $Fe_{1-x}Ni_xPS_3$ | | 760-700 | 690-600 | 9 | 89,92 |
| $Ni_{1-x}Mn_xPS_3$ | | 750-720 | 670-550 | 7-12 | 88,90,153 |
| $Mn_{1-x}Co_xPS_3$ | $P_2S_5$ Flux | 540 | - | 3 | 118 |
| $Fe_{1-x}Co_xPS_3$ | $P_2S_5$ Flux | 580 | - | 4 | 115–117 |
| $Ni_{1-x}Co_xPS_3$ | | 625 | 550 | 21 | 91 |
| $Ni_{1-x}Cr_xPS_3$ | | 750 | 550 | 7 | 178 |
| $Mn_{1-x}Zn_xPS_3$ | | 700-650 | 680-600 | 7 | 78,79,86,152 |
| $Fe_{1-x}Zn_xPS_3$ | | 700 | 650 | 7 | 94 |
| $Fe_{1-x}Cu_xPS_3$ | | 597 | 577 | 15 | 188 |
| $Fe_{1-x}Cd_xPS_3$ | | | | | |
| $Mn_{1-x}Fe_xPSe_3$ | | 750 | 550 | 7 | 123 |
| $(Mn,Fe,Ni,Co)PS_3$ | | 650 | - | 18 | 114 |
| $(V,Mn,Fe,Ni,Co)PS_3$ | | 610 | - | 3 | 114 |
| $(Zn,Mn,Fe,Ni,Co)PS_3$ | | 610 | - | 3 | 114 |
| $(Mg,Mn,Fe,Ni,Co)PS_3$ | | 610 | - | 3 | 114 |
| $(Fe,Mn,Cd,In)PSe_3$ | | 627 | 377 | 7 | 122 |
| $Cu_xNi_{2(1-x)}Cr_xP_2S_6$ | | 750 | 550 | 7 | 182 |
| $MnPS_{3-x}Se_x$ | | 650 | 600 | 7 | 31,32 |
| $NiPS_{3-x}Se_x$ | | 750 | 550 | 7 | 31 |
| $FePS_{3-x}Se_x$ | | 750 | 550 | 7 | 123 |

**Table 2:** Crystal structures of $MPX_3$ compounds.

| Materials | Crystal structure | Space group | Ref. |
|---|---|---|---|
| $VPS_3$ | Monoclinic | $C2/m$ | 71,107 |
| $MnPS_3$ | Monoclinic | $C2/m$ | 31,41,88,105,239 |
| $FePS_3$ | Monoclinic | $C2/m$ | 6,72,89,92,105,108,240 |
| $CoPS_3$ | Monoclinic | $C2/m$ | 109,241 |
| $NiPS_3$ | Monoclinic | $C2/m$ | 31,89,90,92,124 |
| $ZnPS_3$ | Monoclinic | $C2/m$ | 152 |
| $CdPS_3$ | Monoclinic | $C2/m$ | 105 |
| $MgPS_3$ | Monoclinic | $C2/m$ | 125 |
| $PdPS_3$ | Monoclinic | $C2/m$ | 125 |
| $SnPS_3$ | Monoclinic | $C2/m$ | 105,106 |
| $HgPS_3$ | Triclinic | $P\bar{1}$ | 125 |
| $PbPS_3$ | Monoclinic | $P21/c$ | 106 |
| $CrPSe_3$ | Monoclinic | $C2/m$ | 104 |
| $MnPSe_3$ | Rhombohedral | $R\bar{3}$ | 31,242 |
| $FePSe_3$ | Rhombohedral | $R\bar{3}$ | 72,243 |
| $NiPSe_3$ | Monoclinic | $C2/m$ | 125 |
| $ZnPSe_3$ | Rhombohedral | $R\bar{3}$ | 244 |
| $CdPSe_3$ | Rhombohedral | $R\bar{3}$ | 125 |
| $MgPSe_3$ | Rhombohedral | $R\bar{3}$ | 125 |
| $HgPSe_3$ | Monoclinic | $C2/m$ | 245 |
| $PbPSe_3$ | Monoclinic | $P21/c$ | 106 |
| $CuCrP_2S_6$ | Monoclinic | $C2/c$ | 154,155 |
| $CuInP_2S_6$ | Trigonal | $P\bar{3}1c$ | 246 |
| $AgCrP_2S_6$ | Monoclinic | $P2/a$ | 157 |
| $AgInP_2S_6$ | Trigonal | $P31c$ | 247,248 |
| $Mn_{1-x}Fe_xPS_3$ | Monoclinic | $C2/m$ | 82 |
| $Fe_{1-x}Ni_xPS_3$ | Monoclinic | $C2/m$ | 89 |
| $Ni_{1-x}Mn_xPS_3$ | Monoclinic | $C2/m$ | 88,90,153 |
| $Mn_{1-x}Co_xPS_3$ | Monoclinic | $C2/m$ | 118 |
| $Fe_{1-x}Co_xPS_3$ | Monoclinic | $C2/m$ | 115–117 |
| $Ni_{1-x}Co_xPS_3$ | Monoclinic | $C2/m$ | 91 |
| $Ni_{1-x}Cr_xPS_3$ | Monoclinic | $C2/m$ | 178 |
| $Mn_{1-x}Zn_xPS_3$ | Monoclinic | $C2/m$ | 152 |
| $Fe_{1-x}Zn_xPS_3$ | Monoclinic | $C2/m$ | 94 |
| $Mn_{1-x}Fe_xPSe_3$ | Rhombohedral | $R\bar{3}$ | 87 |
| $(Mn,Fe,Ni,Co)PS_3$ | Monoclinic | $C2/m$ | 114 |
| $(V,Mn,Fe,Ni,Co)PS_3$ | Monoclinic | $C2/m$ | 114 |
| $(Zn,Mn,Fe,Ni,Co)PS_3$ | Monoclinic | $C2/m$ | 114 |
| $(Mg,Mn,Fe,Ni,Co)PS_3$ | Monoclinic | $C2/m$ | 114 |
| $(Fe,Mn,Cd,In)PSe_3$ | Rhombohedral | $R\bar{3}$ | 122 |
| $Cu_xNi_{2(1-x)}Cr_xP_2S_6$ | Monoclinic | $C2/m$ | 182 |
| $MnPS_{3-x}Se_x$ | Mono→Rhom | $C2/m \rightarrow R\bar{3}$ | 32 |
| $NiPS_{3-x}Se_x$ | Monoclinic | $C2/m$ | 31 |
| $FePS_{3-x}Se_x$ | Mono→Rhom | $C2/m \rightarrow R\bar{3}$ | 123 |

**Table 3:** Orbital occupancy for 3*d* valence electrons of various transition metal cations under octahedral crystal field.

| Transition metal ions | 3*d* orbital occupancy |
|---|---|
| $V^{2+}$ ($3d^3$) | $e_g$ [↑][↑]<br>$t_{2g}$ [↑][↑][↑] |
| $V^{3+}$ ($3d^2$) | $e_g$ [↑][↑]<br>$t_{2g}$ [↑][↑][↑] |
| $Cr^{2+}$ ($3d^4$) | $e_g$ [↑][↑]<br>$t_{2g}$ [↑][↑][↑] |
| $Cr^{3+}$ ($3d^3$) | $e_g$ [↑][↑]<br>$t_{2g}$ [↑][↑][↑] |
| $Mn^{2+}$($3d^5$) | $e_g$ [↑][↑]<br>$t_{2g}$ [↑][↑][↑] |
| $Fe^{2+}$ ($3d^6$) | $e_g$ [↑][↑]<br>$t_{2g}$ [↑↓][↑][↑] |
| $Fe^{3+}$ ($3d^5$) | $e_g$ [↑][↑]<br>$t_{2g}$ [↑][↑][↑] |
| $Co^{2+}$ ($3d^7$) | $e_g$ [↑][↑]<br>$t_{2g}$ [↑↓][↑↓][↑] |
| $Ni^{2+}$ ($3d^8$) | $e_g$ [↑][↑]<br>$t_{2g}$ [↑↓][↑↓][↑↓] |

**Table 4:** Magnetic exchange and anisotropy parameters for *M*P*X*$_3$ materials. Here, negative and positive signs denote AFM and FM interactions, respectively. $\theta$ is defined as the angle between the *z*-axis and the vector that joins the metal $M^{2+}$ ion and any particular nearest-neighbor chalcogen ligand [Fig. 3(e)]. Without trigonal distortion, $\theta$ becomes 54.7 °.

| Parameters | $VPS_3$*[134] | $MnPS_3$[35] | $FePS_3$[108] | $CoPS_3$[130] | $NiPS_3$[249] | $MnPSe_3$[95] |
|---|---|---|---|---|---|---|
| $S$ | 3/2 | 5/2 | 2 | 3/2 | 1 | 5/2 |
| $T_N$ (K) | 56[107] | 78 | 118 | 120 | 155 | 74 |
| $J_1$ (meV) | -7.387 | −0.77(9) | 1.46(1) | 2.04 | 1.9(1) | -0.45 |
| $J_2$ (meV) | -0.068 | −0.07(7) | −0.04(4) | 0.26 | −0.1(1) | -0.03 |
| $J_3$ (meV) | -0.223 | −0.18(1) | −0.96(5) | -4.21 | −6.90(5) | -0.19 |
| $J_c$ (meV) | - | 0.0019(2) | −0.0073(3) | - | 0.32(3)[131] | -0.031(5) |
| $A$ (meV) | - | 0.0086(9) | 2.66(8) | 2.06 | 0.3(1) | - |
| $\theta$ (°) | - | 51.67[130] | 51.28[130] | 51.38[130] | 51.05[130] | - |

*Theoretical study

**Table 5:** Summary of magnetic properties of high-entropy $MPX_3$ compounds.

| ***HE*P*X*$_3$ compounds** | **Magnetic properties** | **Ref.** |
|---|---|---|
| $Mn_{0.25}Fe_{0.25}Co_{0.25}Ni_{0.27}P_{1.04}S_3$ | AFM ($T_N$ = 70 K) and spin-glass ($T_{g1}$ = 35 K, $T_{g2}$ = 56 K) | 114 |
| $Zn_{0.29}Mn_{0.14}Fe_{0.17}Co_{0.18}Ni_{0.24}PS_{2.61}$ | spin-glass ($T_g$ = 30 K), $T_{kink}$ = 120 K | 114 |
| $Mg_{0.19}Mn_{0.18}Fe_{0.19}Co_{0.26}Ni_{0.28}P_{1.08}S_3$ | Multikinks in MT at 8 K, 42 K, 60, and 120 K | 114 |
| $V_{0.16}Mn_{0.18}Fe_{0.21}Co_{0.23}Ni_{0.24}PS_{2.62}$ | spin-glass ($T_g$ = 37 K), $T_{kink}$ = 150 K | 114 |
| $(FeMnCd)_{0.25}In_{0.17}PSe_3$ | spin-glass ($T_g$ = 15 K) | 122 |

**Table 6:** Summary of magnetic ordering in atomically thin layers of $MPX_3$ compounds.

| Compounds | Magnetic ordering temperature (K) | | | Ref. |
|---|---|---|---|---|
| | Monolayer | Bilayer | Bulk | |
| $MnPS_3$ | -* | 78 | 78 | 29,110 |
| $FePS_3$ | 118 | 118 | 118 | 6 |
| $NiPS_3$ | -* | 130 | 155 | 60 |
| $CoPS_3$ | 100 | 100 | 120 | 109,130 |
| $MnPSe_3$ | 40 | 56 | 74 | 43 |
| $FePSe_3$ | 98 | 102 | 111 | 42 |

*Magnetic order not stable in the monolayer

# Figures

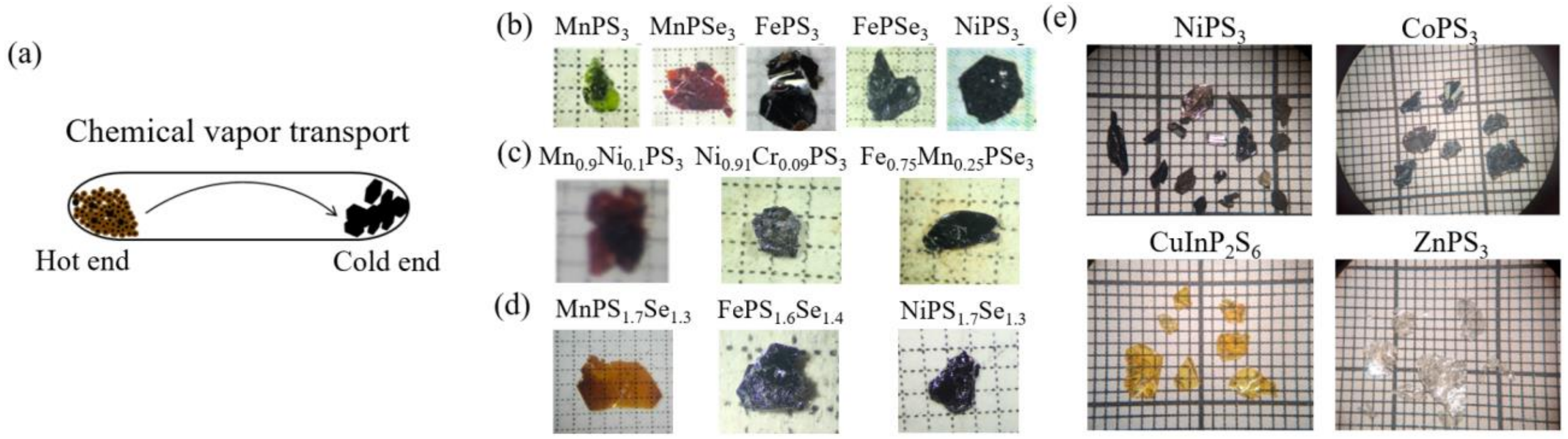

**FIG. 1.** Crystal growth of *M*P*X*3. (a) Schematic of chemical vapor transport (CVT) growth. Optical microscope images of (b) pristine, (c) metal (*M*)-, (d) chalcogen (*X*)-substituted *M*P*X*3 single crystals synthesized by CVT methods. (e) Optical images of single crystals of *M*P*X*3 compounds synthesized by using the reactive $P_2S_5$ flux method. Reprinted (adapted) with permission[116]. Copyright 2021, American Chemical Society.

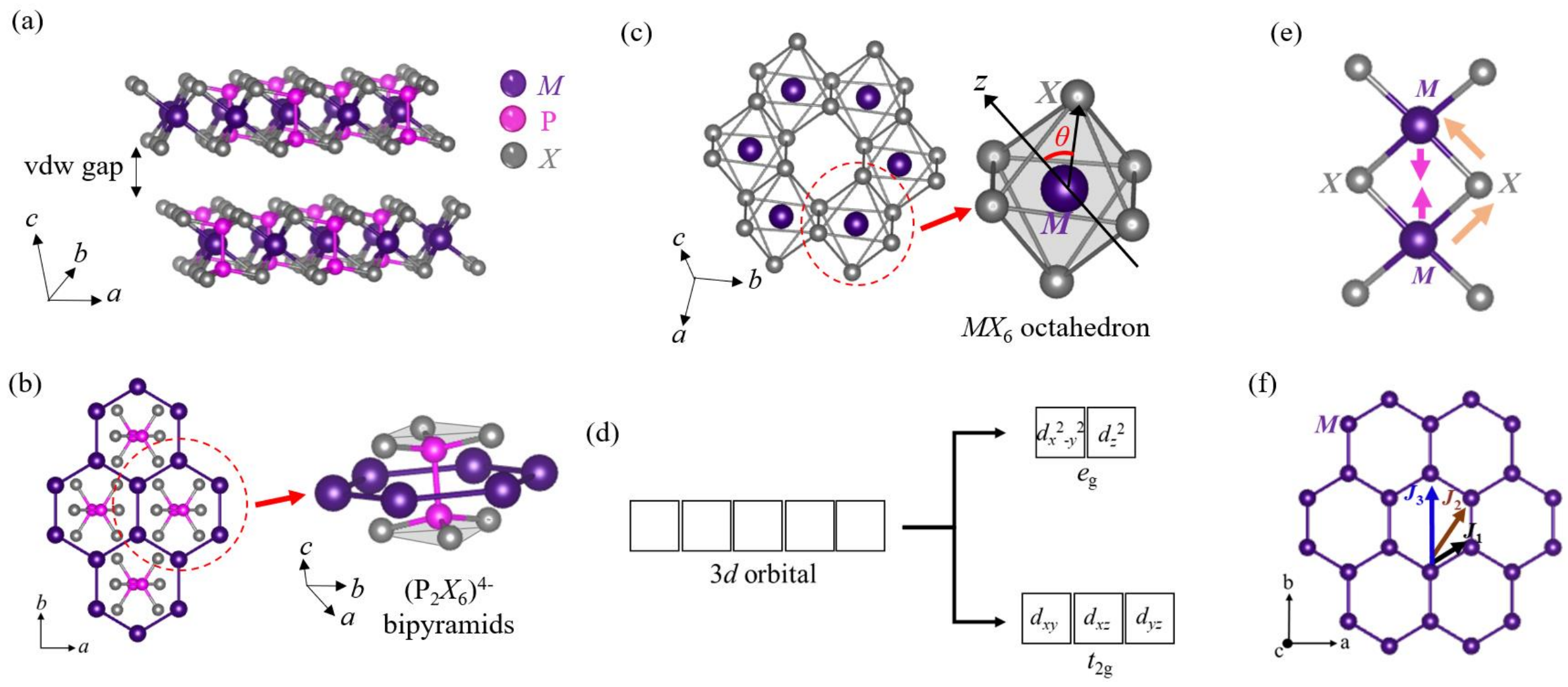

**FIG. 2.** Crystal structures of $MPX_3$. (a) The layered structure of van der Waals material $MPX_3$. (b) The $(P_2X_6)^{4-}$ ($X$ = S or Se) anion sublattice in each layer. Metal $M$ is arranged in a honeycomb lattice around the $(P_2X_6)^{4-}$ bipyramids. (c) $MX_6$ octahedron of $MPX_3$. Such $MX_6$ octahedra possesses a trigonal distortion that is characterized by the angle $\theta$ between the trigonal axis (perpendicular to the *ab*-plane) and the $M$-$X$ bond. (d) Splitting of five 3$d$ orbitals of transition metal cations into three $t_{2g}$ and two $e_g$ levels under an octahedral crystal field. (e) Schematic of direct exchange interaction between magnetic ions, and superexchange interactions between two cations through anion. (f) Schematic of nearest-neighbor ($J_1$), second nearest-neighbor ($J_2$), and third nearest-neighbor ($J_3$) interactions in $MPX_3$.

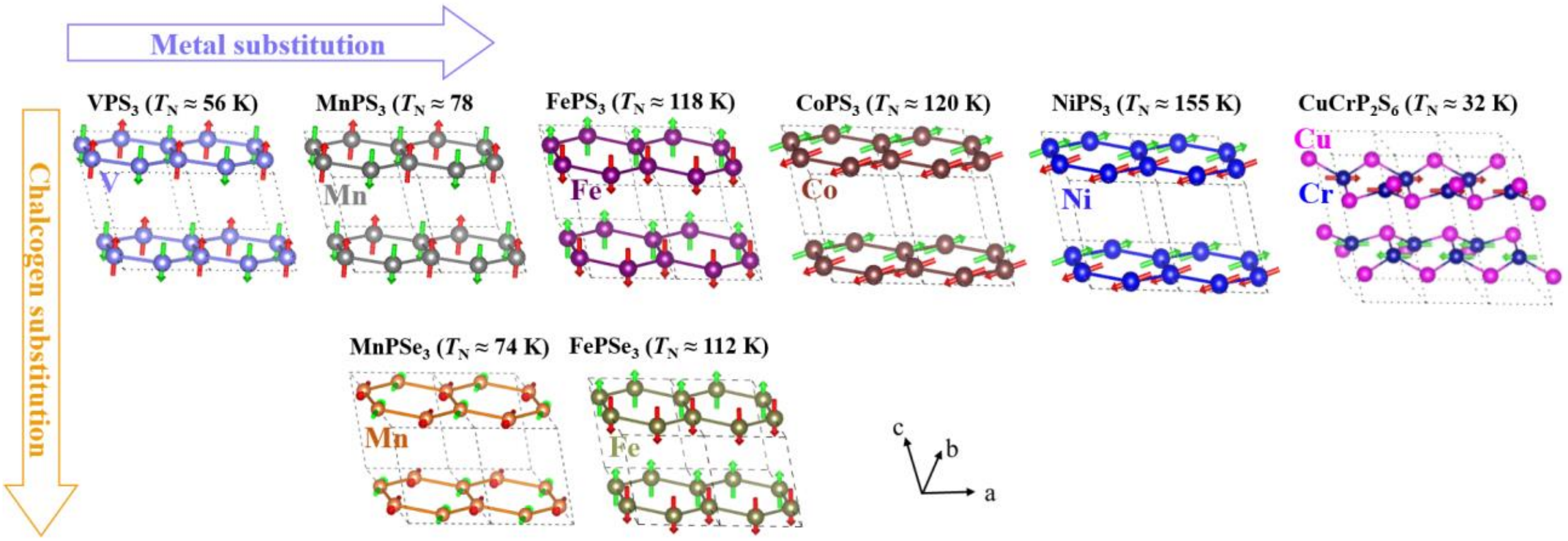

**FIG. 3.** Experimentally determined magnetic structures of various $MPX_3$ materials. For CuCrP2S6, the low-temperature structure is shown. Magnetic structures for (V,Co,Ni)$PSe_3$ and $CuCrP_2Se_6$ have not been experimentally determined.

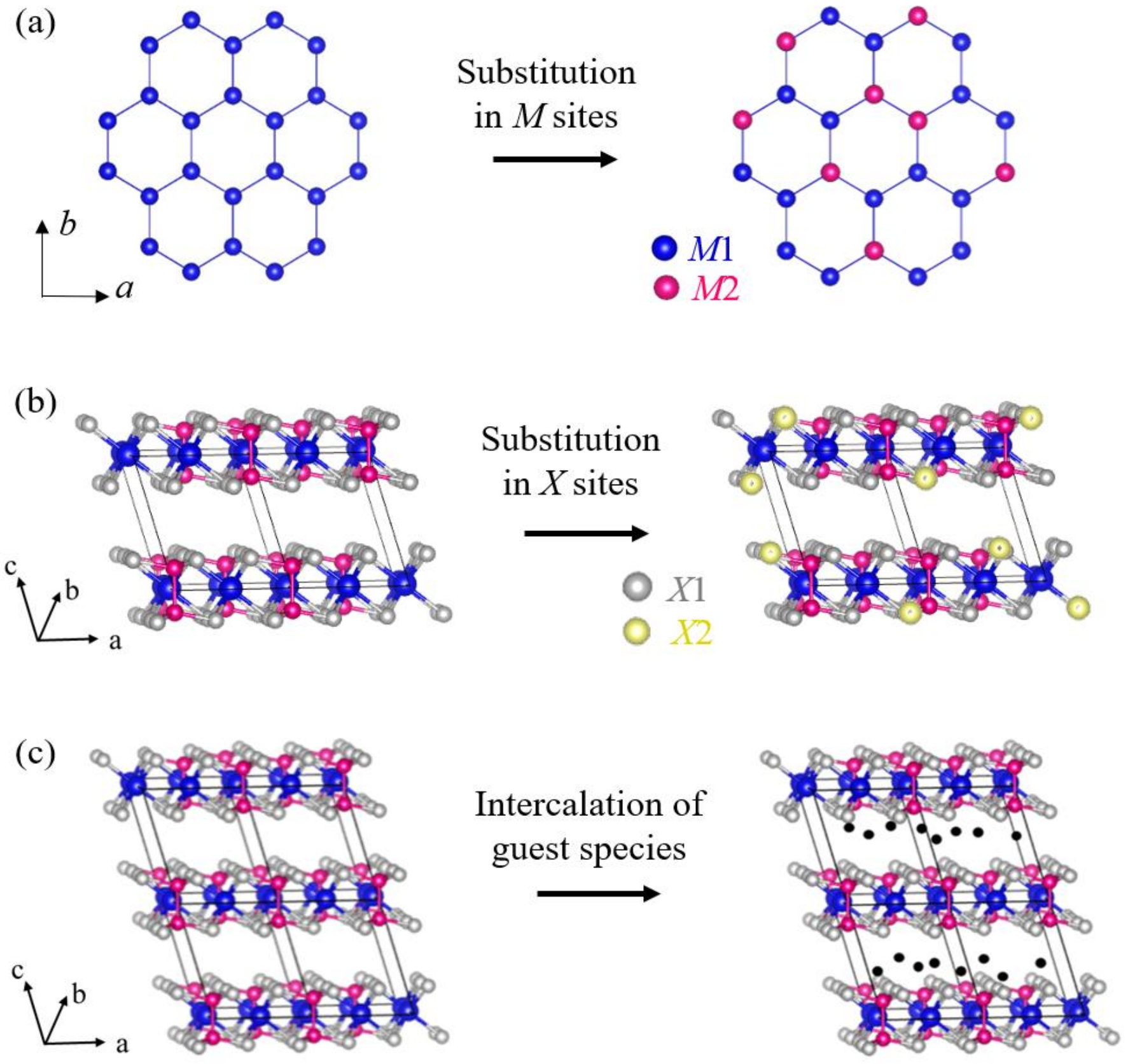

**FIG. 4.** Conceptual schematic of (a) metal (*M*) substitution, (b) chalcogen (*X*) substitution, and (c) Li intercalation in $MPX_3$.

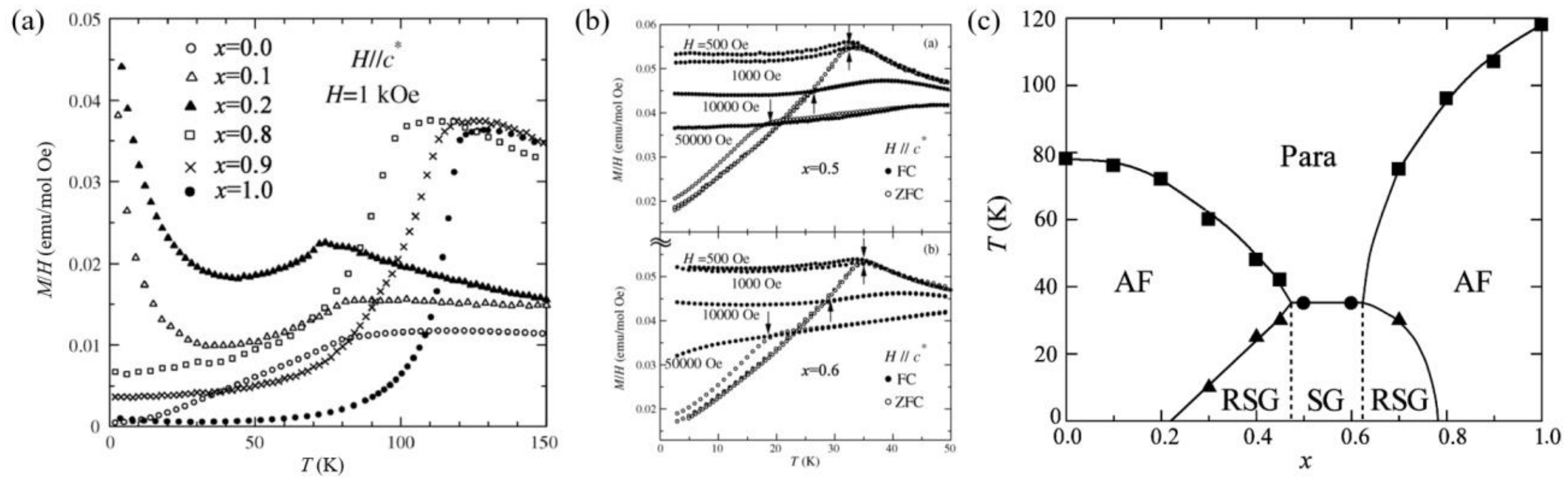

**FIG. 5.** (a-b) Temperature dependence of zero-field cooled (ZFC) susceptibility with the magnetic field applied parallel to the magnetic easy axis ($c^*$-axis) of $Mn_{1-x}Fe_xPS_3$ ($x$ = 0.0, 0.1, 0.2, 0.8, 0.9, 1.0). (c) Magnetic phase diagram of the $Mn_{1-x}Fe_xPS_3$. SG, RSG, AF, and Para indicate a spin glass phase, a reentrant spin glass phase, an antiferromagnetic ordered phase, and a paramagnetic phase, respectively. Reprinted (adapted) with permission[82]. Copyright 2008, Elsevier.

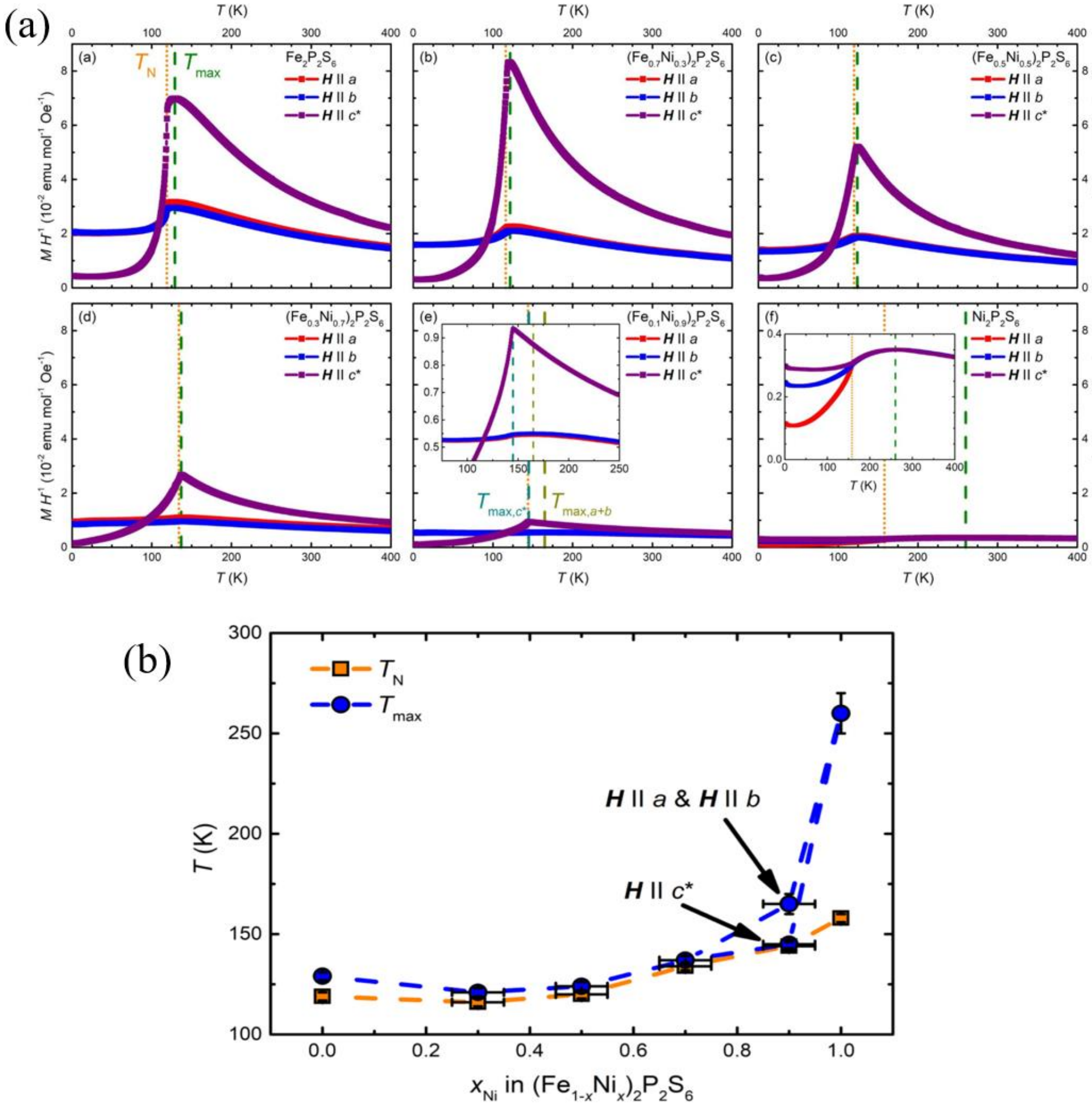

**FIG. 6.** (a) Temperature dependencies of magnetization divided by magnetic field $MH^{-1}$ for $Fe_2P_2S_6$, $(Fe_{0.7}Ni_{0.3})_2P_2S_6$, $(Fe_{0.5}Ni_{0.5})_2P_2S_6$, $(Fe_{0.3}Ni_{0.7})_2P_2S_6$, $(Fe_{0.1}Ni_{0.9})_2P_2S_6$, and $Ni_2P_2S_6$, measured with a field of 1 T applied along three different crystallographic directions. $T_N$ and $T_{max}$ denote Néel temperature and the maximum magnetization temperature. (b) Evolution of $T_N$ and $T_{max}$ with composition for $(Fe_{1-x}Ni_x)_2P_2S_6$. Reprinted (adapted) with permission[89]. Copyright 2021, American Physical Society.

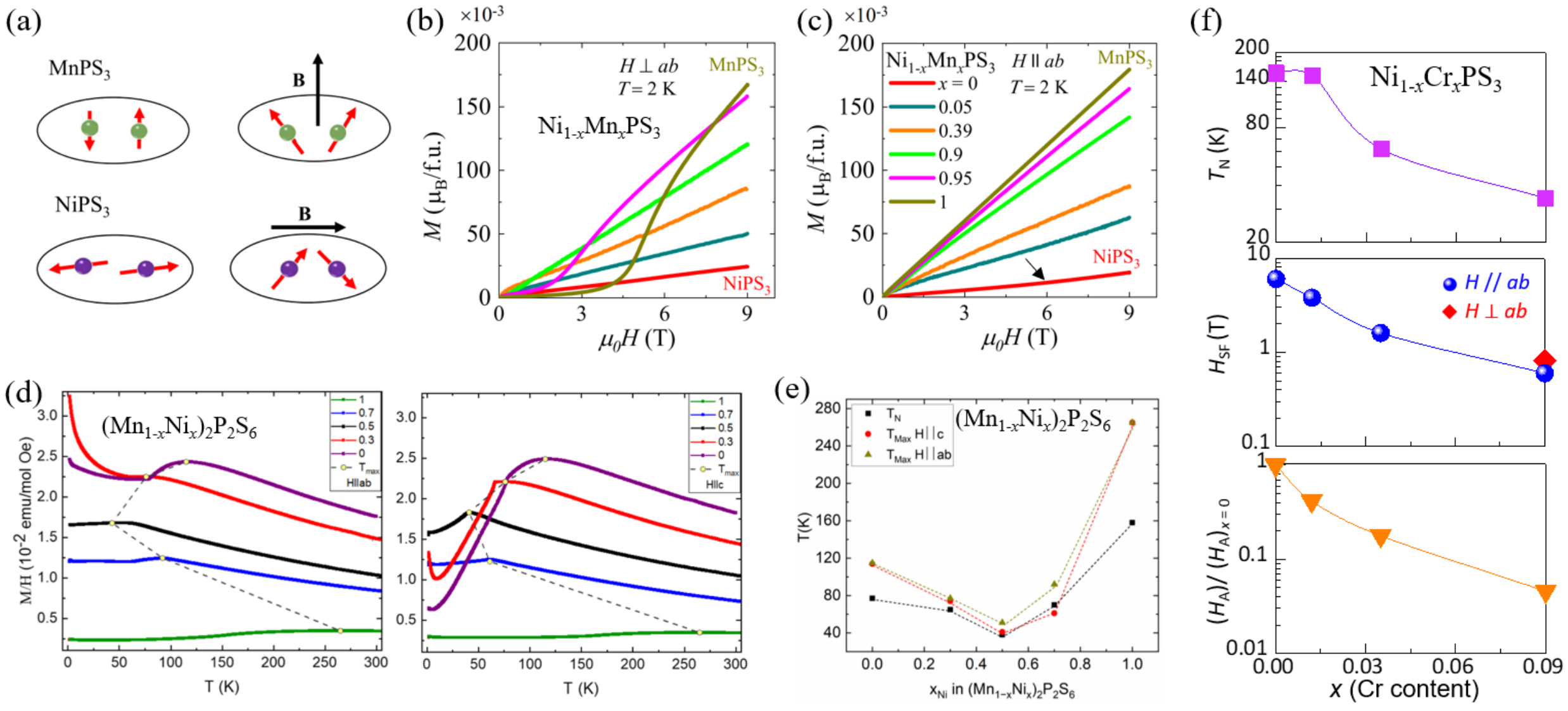

**FIG. 7.** (a) Schematics of spin-flop transition for $MnPS_3$ and $NiPS_3$[90]. (b-c) Magnetic field dependence of magnetization for $Ni_{1-x}Mn_xPS_3$ under in-plane (*H//ab*) and out-of-plane (*H⊥ab*) magnetic fields[90]. Reprinted (adapted) with permission[90]. Copyright 2021, American Physical Society. (d) Temperature dependence of magnetization for $(Mn_{1-x}Ni_x)_2P_2S_6$ under in-plane (*H//ab*) and out-of-plane (*H//c*) magnetic fields[88]. (e) Evolution of $T_N$ and $T_{max}$ with composition for $(Mn_{1-x}Ni_x)_2P_2S_6$[88]. (f) Doping dependence of Néel temperature ($T_N$), spin-flop field ($H_{SF}$), and the effective magnetic anisotropy field ($H_A$) of $Ni_{1-x}Cr_xPS_3$[178].

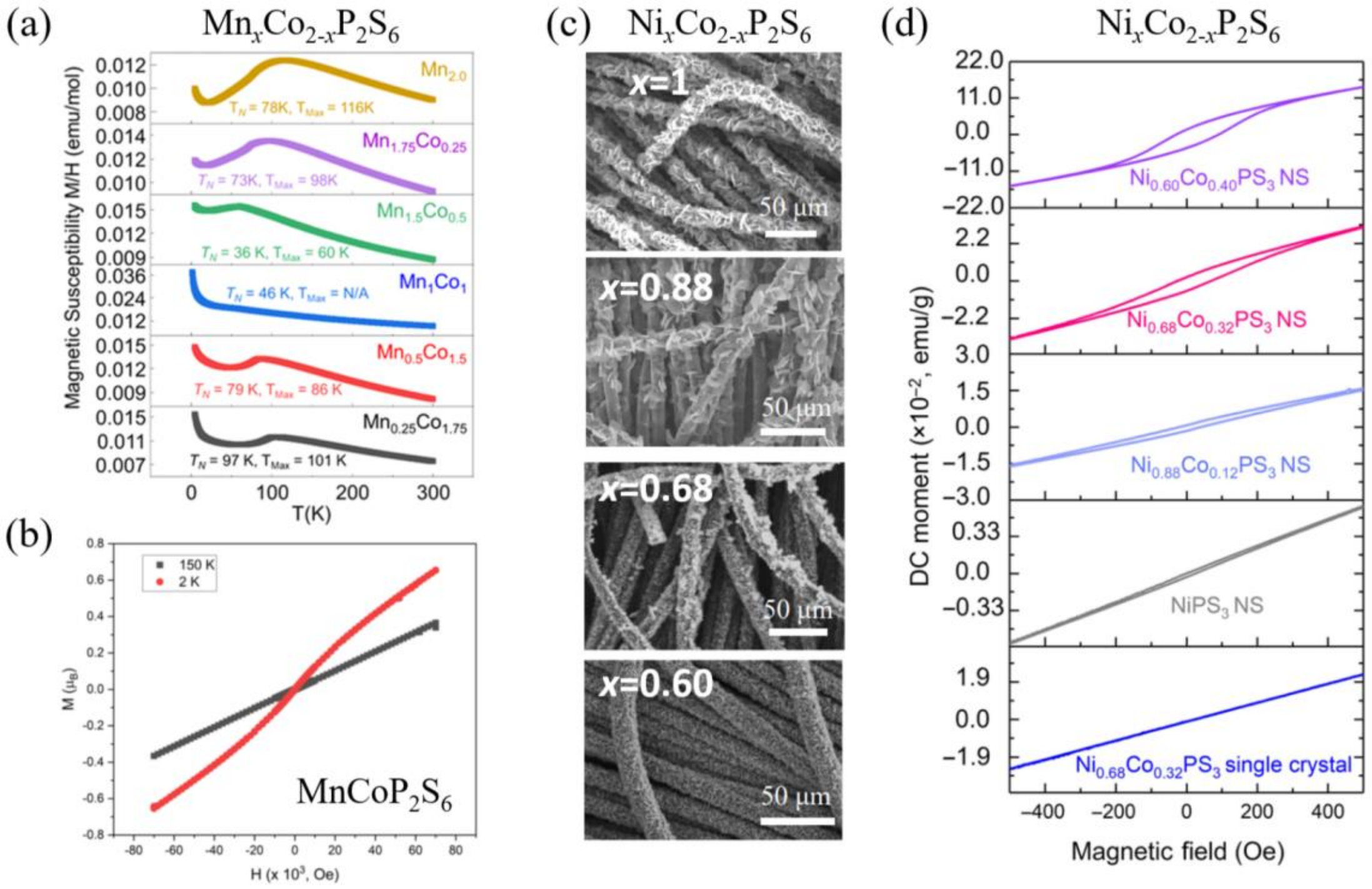

**FIG.8.** (a) Field-cooled (FC) magnetic susceptibility of $Mn_xCo_{2-x}P_2S_6$ measured with $H$ = 1000 Oe[118]. (b) Magnetization for $MnCoP_2S_6$ at 2 K and 150 K[118]. Reprinted (adapted) with permission[118]. Copyright 2022, American Chemical Society. (c) SEM images of nanosheet (NS) of $NiPS_3$, $Ni_{0.88}Co_{0.12}PS_3$, $Ni_{0.68}Co_{0.32}PS_3$, and $Ni_{0.60}Co_{0.40}PS_3$[91]. (d) Magnetization at 5 K for $Ni_{1-x}Co_xPS_3$ NS and single crystal samples[91].

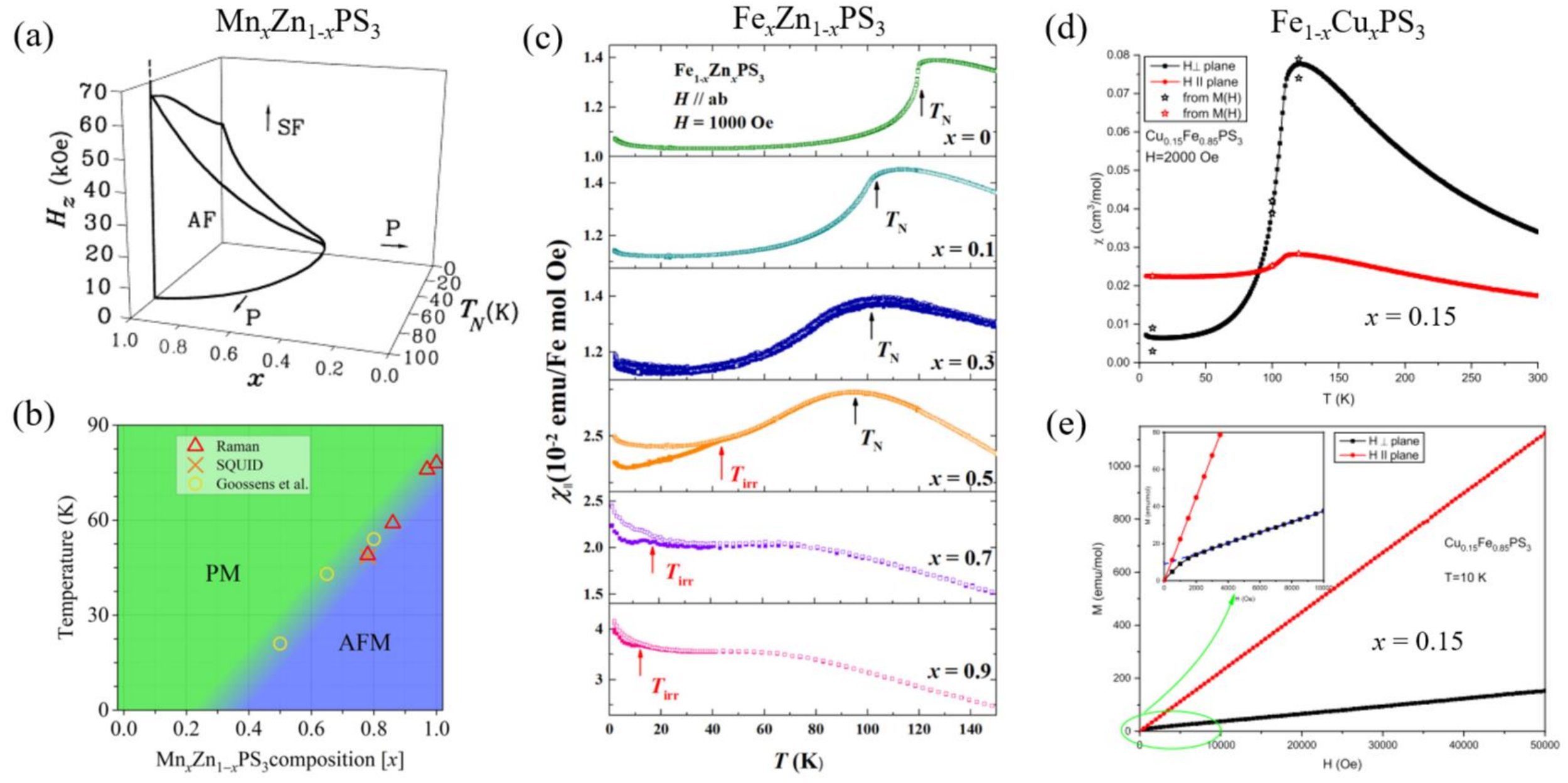

**FIG. 9.** (a-b) Magnetic phase diagrams for $Mn_xZn_{1-x}PS_3$. Field dependence of phases is also shown in (a)[86,152]. Reprinted (adapted) with permission[86]. Copyright 1998, IOP Publishing. Reprinted (adapted) with permission[152]. Copyright 2023, American Physical Society. (c) Temperature dependence of magnetic susceptibilities of $Fe_{1-x}Zn_xPS_3$ under an in-plane magnetic field of 1000 Oe[94]. (d) Temperature dependence of the magnetic susceptibility for $Cu_{0.15}Fe_{0.85}PS_3$ measured under in-plane (H||plane) and out-of-plane (H⊥plane) field of 2000 Oe. (e) Field dependence of magnetization for $Cu_{0.15}Fe_{0.85}PS_3$ at $T$ = 10 K[188].

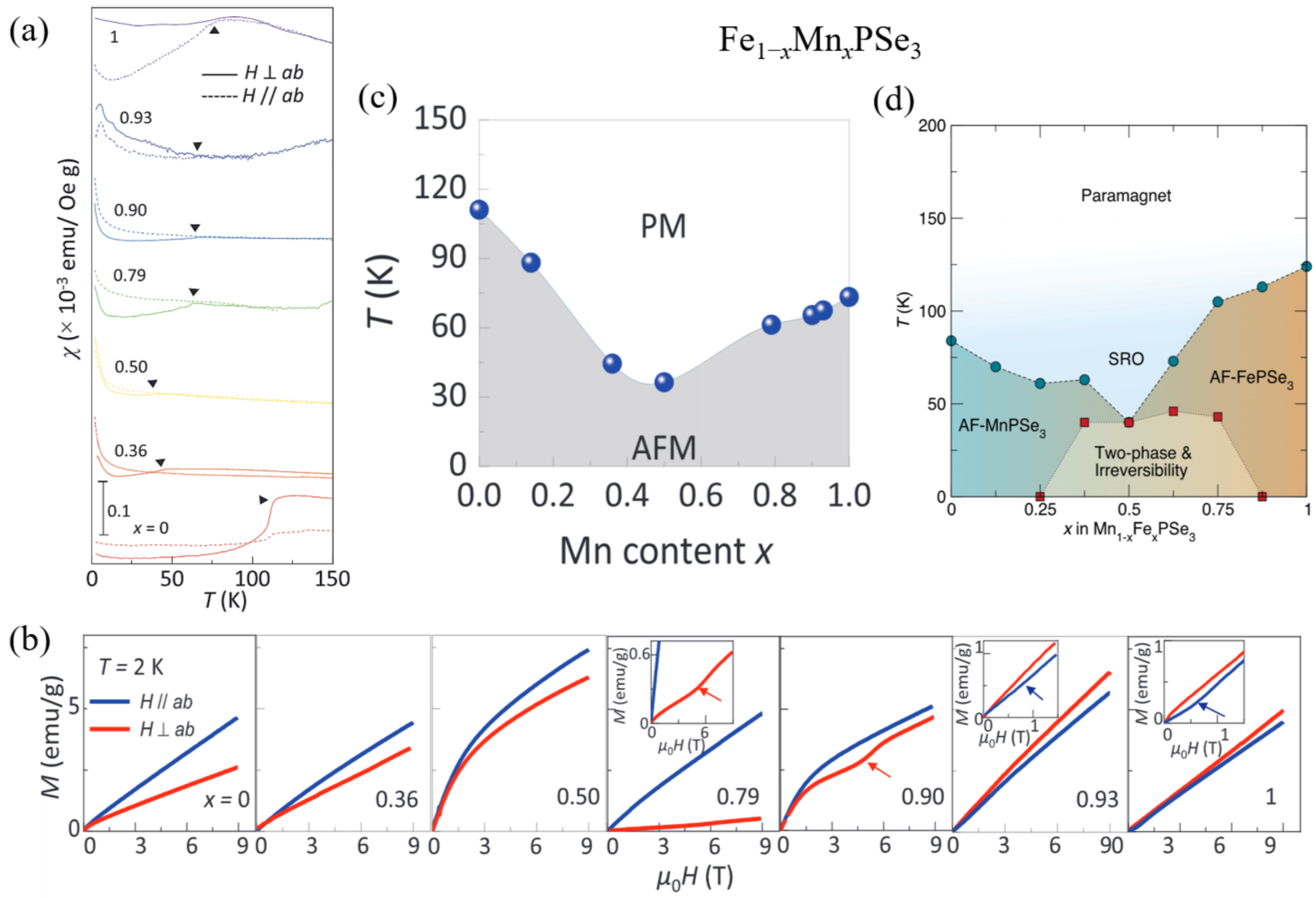

**FIG. 10.** (a) Temperature dependencies susceptibility for $Fe_{1-x}Mn_xPSe_3$ measured under a magnetic field of 0.1 T[123]. (b) Isothermal magnetization at 2 K for $Fe_{1-x}Mn_xPSe_3$ under out-of-plane ($H \perp ab$, red) and in-plane ($H \| ab$, blue) magnetic fields. Insets show spin-flop transition at low fields[123]. (c-d) Magnetic phase diagrams of $Mn_{1-x}Fe_xPSe_3$ obtained from (c) single crystals[123] and (d) polycrystals[87]. Reprinted (adapted) with permission[87]. Copyright 2020, American Physical Society.

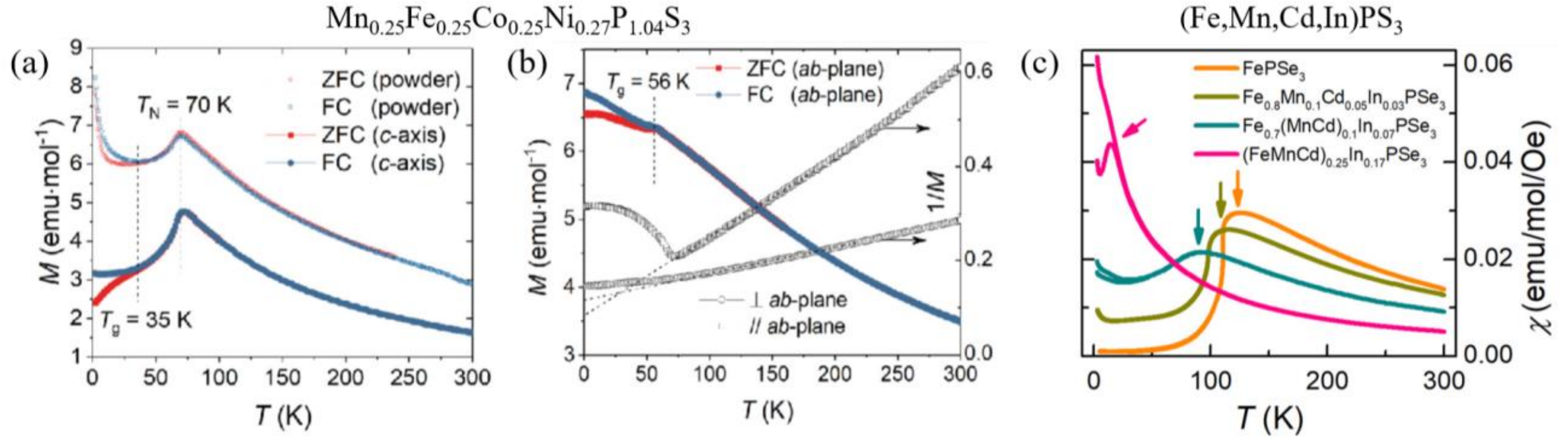

**FIG. 11.** (a-b) Magnetization of (Mn,Fe,Co,Ni)$PS_3$ under the external magnetic field of 500 Oe along different directions. The antiferromagnetic transition ($T_N$) and spin glass transition temperature ($T_g$) are indicated[114]. Reprinted (adapted) with permission[114]. Copyright 2021, American Chemical Society. (c) Susceptibility ($\chi$) of pristine $FePSe_3$ and (Fe,Mn,Cd,In)$PS_3$[122]. Reprinted (adapted) with permission[122]. Copyright 2022, American Physical Society.

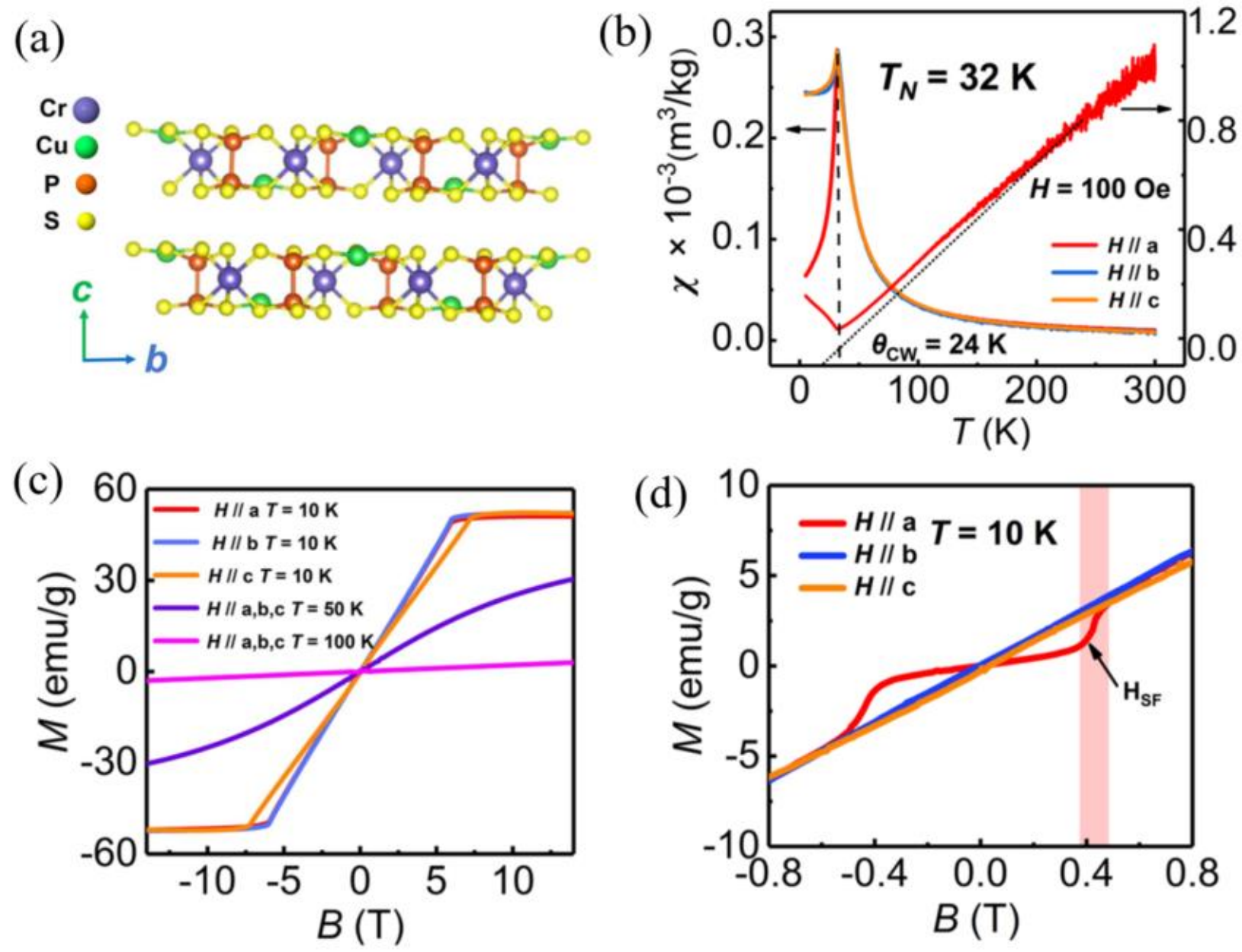

**FIG. 12.** (a) Crystal structure of $CuCrP_2S_6$. (b-c) Temperature and magnetic field dependencies of magnetic susceptibility of $CuCrP_2S_6$. (d) Spin-flop transition when the field is applied along the a-axis[155].

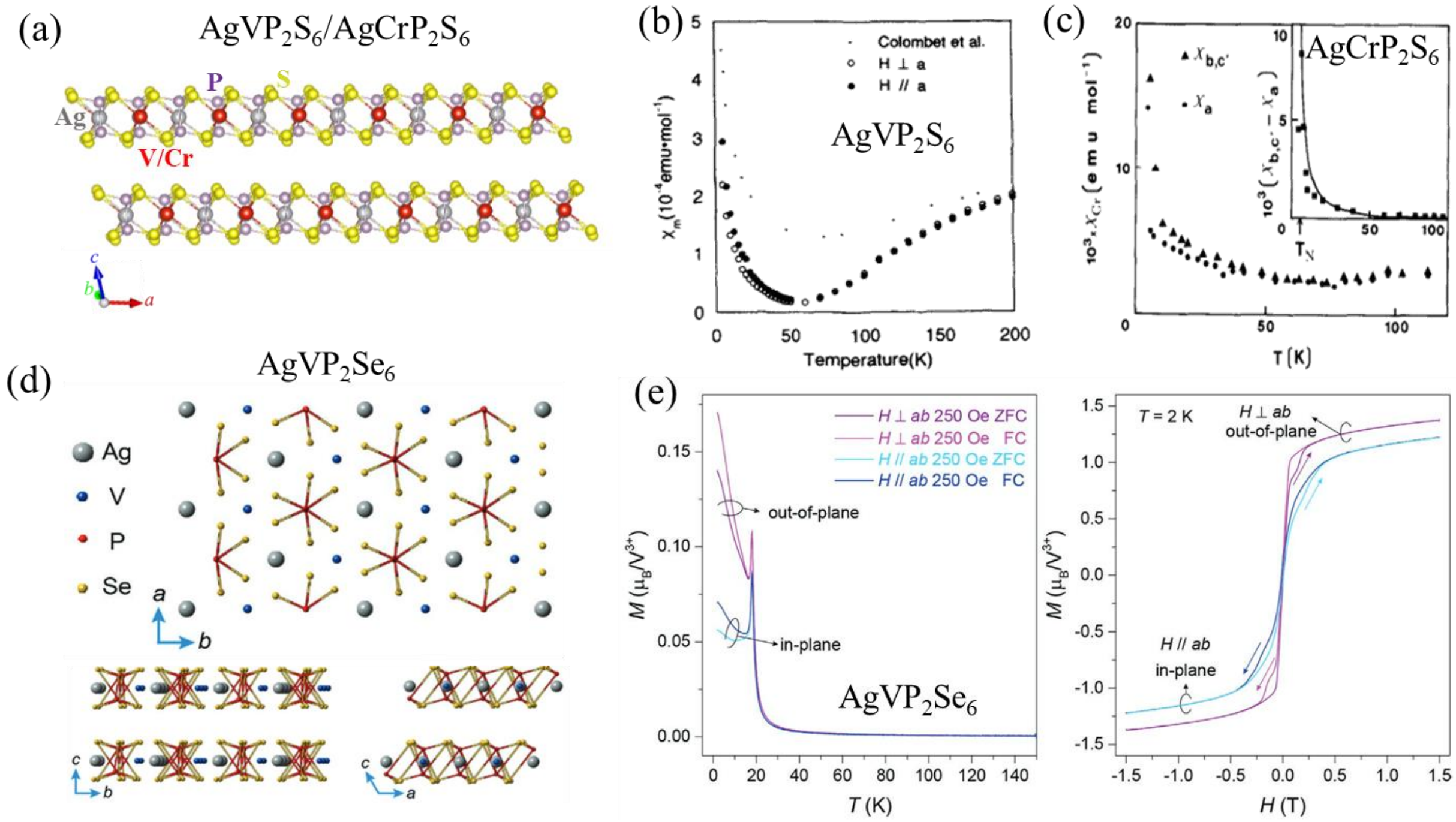

**FIG. 13.** (a) Crystal structure of $AgVP_2S_6$ and $AgCrP_2S_6$. (b-c) Temperature dependence of the susceptibility for $AgVP_2S_6$[192] and $AgCrP_2S_6$[193]. Reprinted (adapted) with permission[192]. Copyright 1994, Elsevier. Reprinted (adapted) with permission[193]. Copyright 1990, Elsevier. (d) Crystal structure (top and side views) of $AgVP_2Se_6$. (e) Temperature (left) and magnetic field (right) dependencies of magnetization for $AgVP_2Se_6$[198].

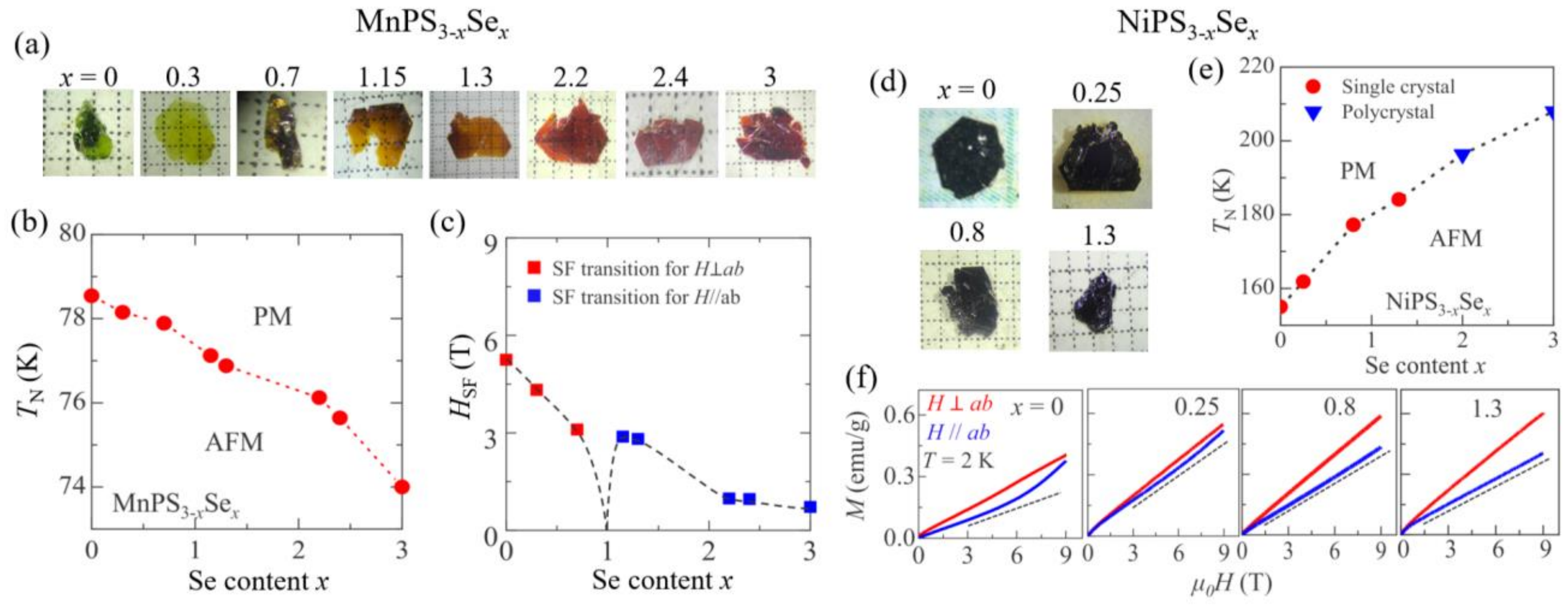

**FIG. 14.** (a) Images of MnPS$_{3-x}$Se$_x$ single crystals. Doping dependence of (b) Néel temperature ($T_N$) and (c) spin-flop (SF) field for MnPS$_{3-x}$Se$_x$. (d) Images of NiPS$_{3-x}$Se$_x$ single crystals ($0 \leq x \leq 1.3$). (e) Doping dependence of $T_N$ for NiPS$_{3-x}$Se$_x$. (f) Field dependence of magnetization of NiPS$_{3-x}$Se$_x$ at $T$ = 2 K[31]. Reprinted (adapted) with permission[31]. Copyright 2022, American Physical Society.

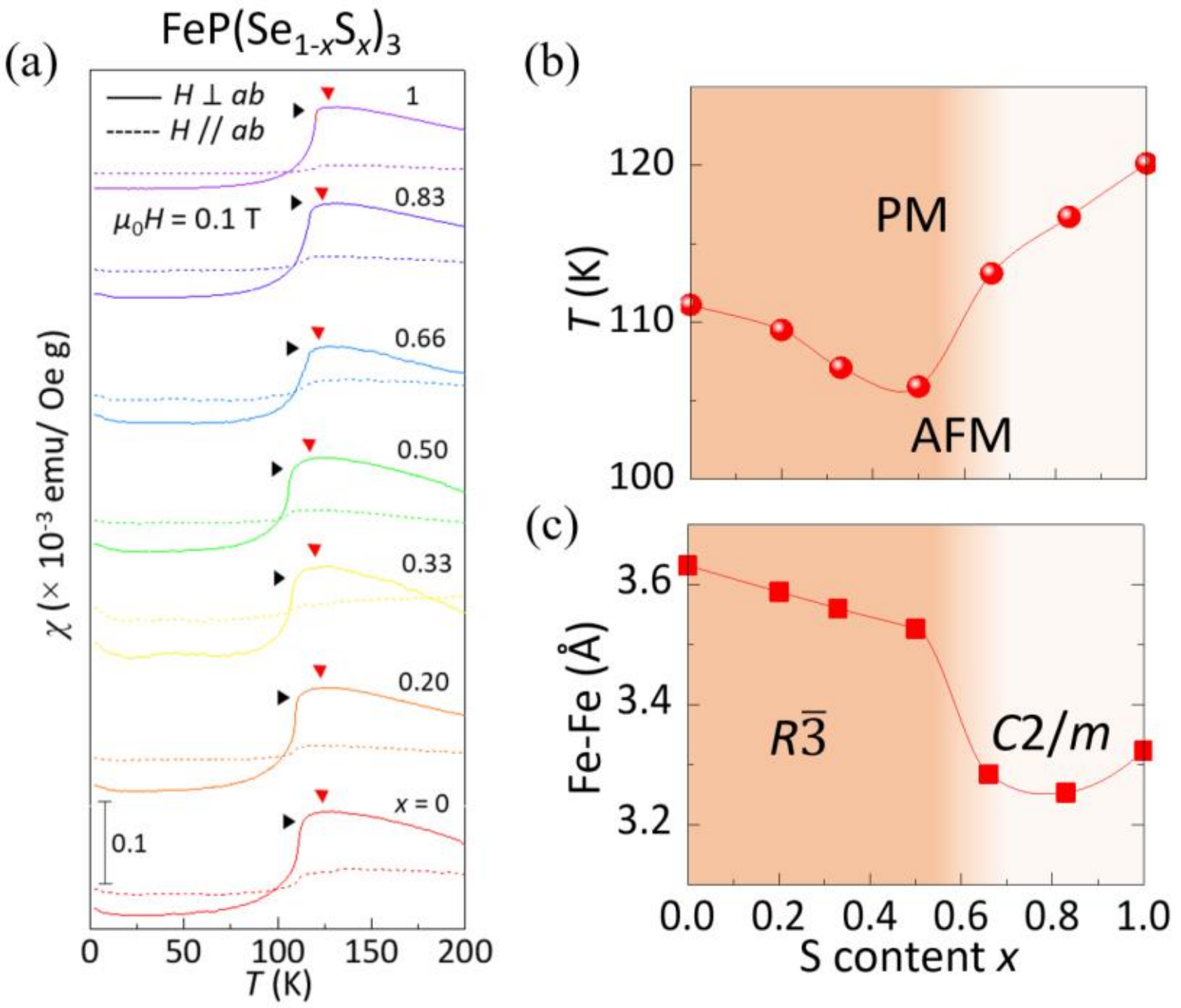

**FIG. 15.** (a) Temperature dependencies of susceptibility of $FeP(Se_{1-x}S_x)_3$ measured under a magnetic field of 0.1 T. Doping dependence of (b) $T_N$ and (c) Fe-Fe distance for $FeP(Se_{1-x}S_x)_3$[123].

(a)

| composition | $\mu_{eff}$, $\mu_B$ | Θ, K | $T_N$, K |
|---|---|---|---|
| $MnPSe_3$ | 6.1 | −201 | 85 |
| $Li_{0.2}MnPSe_3$ | 6.2 | −201 | 87 |
| $Li_{0.5}MnPSe_3$ | 6.1 | −194 | 83 |
| $Li_1MnPSe_3$ | 6.0 | −190 | 85 |
| $FePSe_3$ | 5.0 (5.1) | −4 (37) | 118 (123) |
| $Li_{0.44}FePSe_3$ | 4.9 | 4 | 123 |
| $Li_{0.74}FePSe_3$ | 4.9 | 1 | 118 |
| $FePS_3$ | 5.0 (5.4) | 65 (14) | 133 (126) |
| $Li_{0.5}FePS_3$ | 5.2 | 81 | 153 |
| $Li_{1.2}FePS_3$ | 5.1 | 92 | 135 |
| $NiPS_3$ | 3.9 (3.7) | −712 (−558) | 300 (254) |

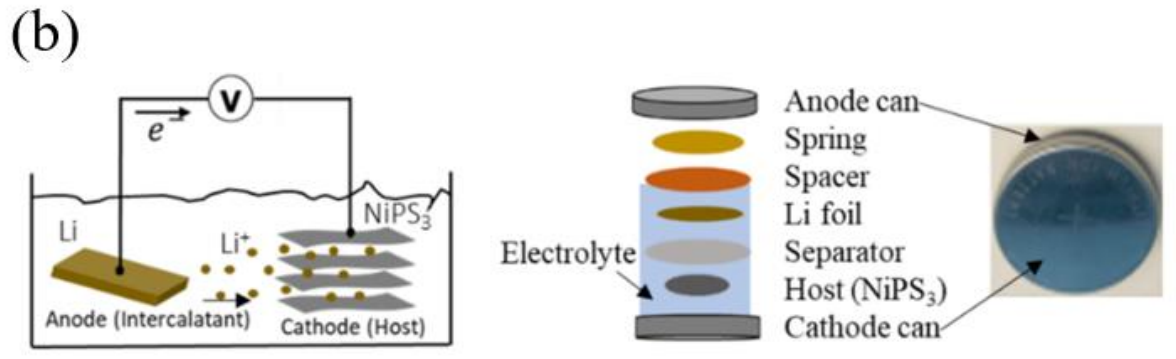

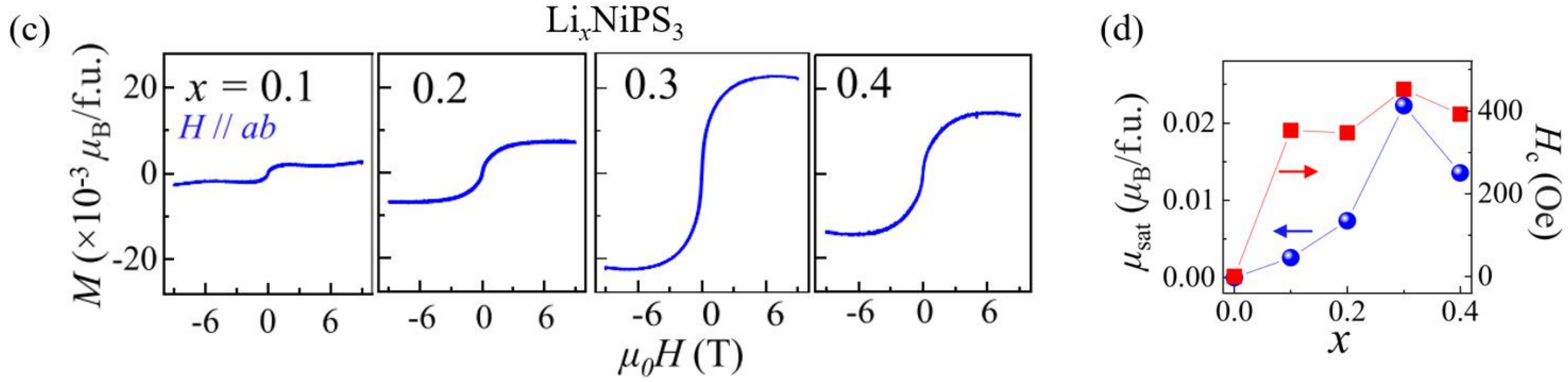

**FIG. 16.** (a) Magnetic properties for $Li_xMnPSe_3$, $Li_xFePSe_3$, $Li_xFePS_3$, and $Li_xNiPS_3$[101]. Reprinted (adapted) with permission[101]. Copyright 1979, American Chemical Society. (b) Conceptual schematic of electrochemical intercalation process and the coin cell setup for intercalation[99]. (c) Non-linear field dependence of magnetization of $Li_xNiPS_3$ at $T$ = 2 K after removing the linear magnetization background[99]. (d) Evolution of saturation moment ($\mu_{sat}$) and coercive fields ($H_c$) with Li content[99].

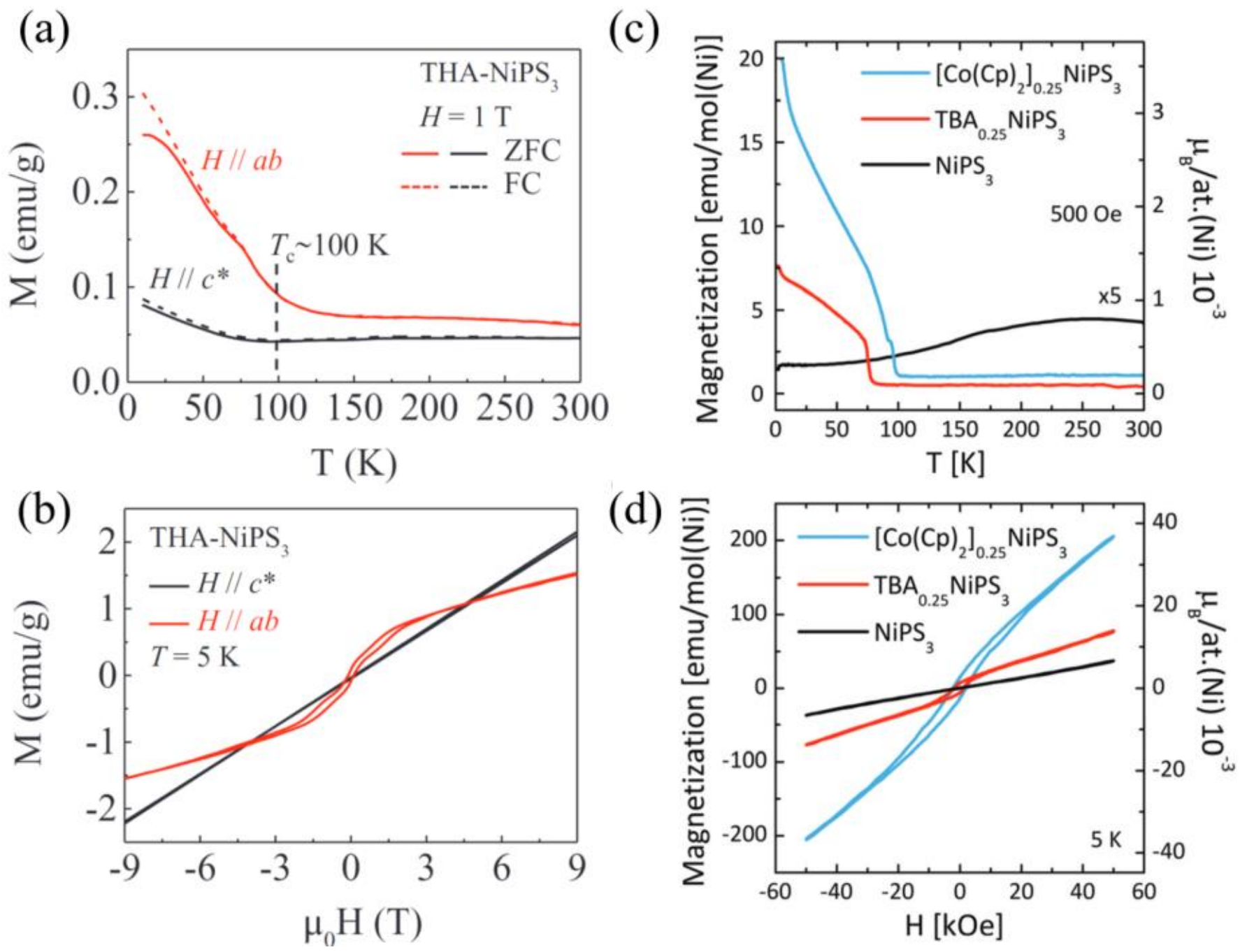

**FIG. 17.** (a) Temperature and (b) field dependencies of magnetization of intercalated THA-$NiPS_3$[98]. (c) Field-cooled temperature-dependent magnetization and (d) magnetic hysteresis for $NiPS_3$, $TBA_{0.25}NiPS_3$, and $[Co(Cp)_2]_{0.25}NiPS_3$[100].